\documentclass{emulateapj}
\usepackage{psfig}
\usepackage{here}
\slugcomment{Submitted to Astrophysical Journal}

\newcommand{\beq} {\begin{equation}} \newcommand{\eeq}
{\end{equation}} \newcommand{\beqa} {\begin{eqnarray}}
\newcommand{\eeqa} {\end{eqnarray}}

\newcommand{\rct} {{\tilde R_c}} 

\begin{document}


\lefthead{Chakrabarti \& Whitney}
\righthead{}

\title{Panchromatic Spectral Energy Distributions of Dusty Galaxies with RADISHE. I. Predictions for $\it{Herschel}$: Correlating Colors With Galactic Energy Sources}
\author{Sukanya Chakrabarti\altaffilmark{1,2} \& Barbara A. Whitney\altaffilmark{3}}

\altaffiltext{1}{
Harvard-Smithsonian Center for Astrophysics, 60 Garden Street, Cambridge, MA 02138 USA, schakrabarti@cfa.harvard.edu}
\altaffiltext{2} {National Science Foundation Postdoctoral Fellow}
\altaffiltext{3}{Space Science Institute, 4750 Walnut St. Suite 205, Boulder, CO 80301, bwhitney@spacescience.org}

\begin{abstract}

We present three-dimensional, self-consistent radiative transfer solutions with a new Monte Carlo radiative equilibrium code.  The code, RADISHE ($\bf{RAD}$iative transfer $\bf{I}$n $\bf{S}$moothed particle $\bf{H}$ydrodynamics and $\bf{E}$ulerian codes), can be applied to calculate the emergent spectral energy distributions (SEDs) and broadband images from optical to millimeter wavelengths of arbitrary density geometries with distributed sources of radiation.  One of the primary uses of this code has been to interface with hydrodynamical codes to calculate emergent SEDs along a simulation time sequence.  We focus the applications of this code in this paper on infrared bright dusty galaxies, but RADISHE is also ideal for calculating emission from star clusters or protostellar environments.  The primary methodological focus of this paper is on the radiative equilibrium temperature calculation.  We find that an iterative calculation of the temperature, which takes as the Monte Carlo estimator for the mean free intensity the sum of photon flight paths, is significantly faster than relaxation temperature calculation methods, particularly when large numbers of grid cells are required, i.e., in modeling three-dimensional geometries such as the dust envelopes of turbulent massive protostellar cores or infrared bright galaxies.  An accurate long wavelength SED and corresponding temperature calculation will be essential for analyzing upcoming $\it{Herschel~Space~Observatory}$ observations.  We present simulated color-color plots for infrared bright galaxies at a range of redshifts, and unfold these plots as color vs the fractional AGN luminosity, to demonstrate that $\it{Herschel}$ will be able to effectively discriminate between submillimeter galaxies where the energy source is dominated by AGN and those where star formation dominates.  We demarcate in particular the ``Class II'' or energetically active AGN evolutionary phase in $\it{Herschel}$ color-color plots.

\end{abstract}



\section{Introduction}

Ultraluminous Infrared Galaxies (ULIRGs) radiate more than $1\times 10^{12}L_{\odot}$ between $8-1000~\micron$ (Rieke \& Low 1972; Soifer et al. 1984; Aaronson et al. 1984; Sanders \& Mirabel 1996).   Submillimeter galaxies (SMGs), a high redshift population of dusty, infrared-bright galaxies, ($F_{850~\micron} \ga 1~\rm mJy$ (Smail et al. 1997; Blain et al. 2002), of which Chapman et al. (2005) were able to obtain spectroscopic redshifts for, guided by optical and radio associations, have been studied with SCUBA (Smail et al. 1997; 2002; Ivison et al. 2000; Ivison et al. 2007), more recently with $\it{Spitzer}$ (Pope et al. 2006), with $\it{Chandra}$ (Alexander et al. 2005), and SHARC-2 (Kovacs et al. 2006).  These observations reveal that SMGs are a cosmologically significant (Blain et al. 2002), diverse (Ivison et al. 2000) population of galaxies, with some SMGs harboring obscured AGN (Alexander et al. 2005; Pope et al. 2006).  The Kovacs et al. (2006) sample of $350~\micron$ observations is currently the most direct probe of the rest-frame far-IR of the SMG population at $z \sim 2$.  Aside from this set of observations, there are few rest-frame far-IR observations of $z \sim 2$ SMGs.  As such, the bolometric luminosities of this population remain uncertain for the majority of this population, as the far-IR luminosity is comparable to the bolometric luminosity.  The $\it{Herschel~Space~Observatory}$ is expected to secure rest-frame far-IR observations of high redshift infrared bright galaxies, which combined with existing data sets can provide for the first time a panchromatic view of the dust emission from this population.

A new development in the last several years is the use of Smoothed Particle Hydrodynamics (SPH) and Eulerian codes to routinely calculate the time evolution of the gaseous and collionless components of galaxies undergoing mergers or evolving quiescently (Kravtsov 2003; Chakrabarti et al. 2003; Springel et al. 2005; Governato et al. 2007; Cox et al. 2006; Kazantzidis et al. 2007).  Calculating the emergent SEDs and images along a time sequence of simulations of merging galaxies using a self-consistent radiative transfer code can allow us to build an interface between simulations of galaxy mergers and observations of dusty galaxies like ULIRGs.  To aid in the interpretation of current and upcoming infrared observations, we have developed a self-consistent, three-dimensional Monte Carlo radiative transfer code.  It is nicknamed $\bf{RADISHE}$ (RADiative transfer In Smooth particle Hydrodynamics and Eulerian codes), as it has been primarily employed to interface with the time outputs of hydrodynamical simulations.  Chakrabarti et al. (2007a;b) employed RADISHE to calculate the emergent SEDs of galaxy mergers.  In particular, Chakrabarti et al. (2007a) used an early version of RADISHE to demonstrate that the cold-warm IRAS classfication of galaxies (de Grijp et al. 1985) can be reproduced by galaxy merger simulations and can be explained due to AGN feedback.  Chakrabarti et al. (2007b) demonstrated that trends in observed color-color plots can be explained by decomposing simulated color-color plots as a function of time or some intrinsic galaxy parameter, such as the ratio of the black hole to stellar bolometric luminosity.  Chakrabarti et al. (2007b) also proposed an evolutionary model for SMGs, where they evolve from high ratios of $L_{\rm IR}/L_{\rm x}$ (Class I) through a bright x-ray bright phase (Class II), to low values of $L_{\rm IR}/L_{\rm x}$ typical of merger remnants (Class III).

The Monte Carlo method is ideal for treating inhomogeneous density distributions where there is no clear symmetry.  Boundary conditions, which must be dealt with explicitly in finite-difference codes (Steinacker et al. 2003), and which often become difficult to treat in multi-dimensional finite-difference codes, are naturally accomodated in the Monte Carlo method as long as the photon statistics are robust.  Analytic methods are essential to reduce the solution of the radiative transfer problem to a few key parameters under simplifying assumptions (Chakrabarti \& McKee 2005; Chakrabarti \& McKee 2007), while numerical methods are needed to model the effects of inhomogeneous geometry and dust grain distributions so as to calculate emergent SEDs in a panchromatic manner.  The turbulent envelopes of massive protostars and disordered morphologies of merging galaxies cannot be modeled in detail by treating them as purely axisymmetric systems, although this is a useful first approximation (Efstathiou \& Rowan-Robinson 1995; Silva et al. 1998).  The Monte Carlo method relies on sampling probability distributions, for which it is necessary to use large numbers of photons.  Although this has been previously computationally restrictive, advances in computing speed and memory, along with the development of new algorithms and codes, have allowed this method to mature into a sophisticated and rigorous means of solving the radiative transfer problem.   Self-consistent Monte Carlo codes have been used to model the dust emission from protostellar regions (Whitney et al. 2003a;200b; Indebetouw et al.2006; Robitaille et al. 2006), to interface with SPH density fields (Stamatellos \& Whitworth 2005), to study protoplanetary disk evolution (Wood et al. 2002), to compute SEDs and images from arbitrary density and stellar distributions (Misselt et al. 2001), 
 and to compute the SEDs of clumpy toroidal models of AGN (Dullemond \& van Bemmel 2005).  The Monte Carlo method has also been used recently to study the effects of dust attenuation and scattering on the emergent UV-near-IR SED from galaxy merger simulations (Jonsson 2006), and more recently to self-consistently compute the multi-wavelength emergent SEDs, images, and photometric time history of merger simulations to explain trends (cold-warm IRAS classfication; IRAC color-color plots) in observed data (Chakrabarti et al. 2007a; 2007b), and to calculate the demographics of infrared bright galaxies locally and as a function of redshift (Chakrabarti, Huang et al., in preparation). 

In this paper, we focus our attention on algorithms designed to enforce radiative equilibrium in Monte Carlo codes.  The radiative equilibrium condition is the self-consistency condition that ensures that the flux of radiation is divergence free.  This is the basic condition that needs to be satisified to calculate the global radiation field and equilibrium dust temperature within the envelope.  Solving for the dust temperature can be accomplished locally only when the envelope is optically thin to its own reprocessed radiation.  There are two general classes of algorithms that have been used to enforce radiative equilibrium in Monte Carlo codes, namely what we refer to as relaxation methods (Bjorkman \& Wood 2001) and iterative methods (Lucy 1999).  We review the details of these methods in \S 2.

The organization of the paper is as follows.  In \S 2, we review two basic types of algorithms designed to enforce radiative equilibrium in Monte Carlo codes.  
In \S 3 we contrast the efficiency of these algorithms for calculating the SEDs and temperatures of homogeneous dust envelopes in one and three-dimensional grids.   We discuss two separate conditions that users may enforce, namely, the number of photons required for the temperature to converge ($N_{\rm temp}$) and for noise requirements on the SED ($N_{\rm SED}$).  In \S 3.3, we present radiative transfer solutions for three-dimensional inhomogeneous envelopes and also compare the emergent SEDs to those of equivalent homogeneous envelopes.  In \S 4, we present the basic setup for RADISHE, convergence studies, and comparisons of our calculated SEDs to observed SEDs of $z \sim 2$ SMGs.  We show in \S 5 that Herschel color-color plots will be able to discern the energy sources of SMGs.  We conclude in \S 6.

\section{The Radiative Equilibrium Temperature}

The emergent radiation from a dusty envelope is a result of three fundamental processes:  attenuation, scattering, and reemission.  Attenuation refers to the absorption of an incident photon from a source of radiation along its trajectory, with the net result that the incident intensity is lowered by $exp(-\tau_{\nu})$ along the trajectory, where $\tau_{\nu}=\int \kappa_{\nu,\rm ext} \rho ds$ is the optical depth along the trajectory, where $\kappa_{\nu,\rm ext}=\kappa_{\nu}+\kappa_{\nu,\rm scat}$, i.e., the total extinction ($\kappa_{\nu,\rm ext}$) is the sum of the absorption ($\kappa_{\nu}$) and scattering ($\kappa_{\nu,\rm scat}$) opacities, and $\rho$ the density along the path length $ds$.  Scattering of photons off dust grains has been treated in detail by previous workers (Witt 1977a;b;c; Whitney \& Hartman 1992; Whitney \& Hartman 1993; Code \& Whitney 1995; Wood et al. 1996a; 1996b; Wood 1998;  Gordon et al. 2001; 
Watson \& Henney 2001;
Whitney \& Wolff 2002).  The implementation of scattering in RADISHE uses a modified Henyey-Greenstein function for the scattering phase function and is the same as that of Whitney et al. (2003); the efficiency of scattering is wavelength-dependent and is proportional to $1/\lambda^{4}$, i.e., scattering is particularly relevant for short wavelengths.  

Reemission refers to the emission of photons from dust grains subsequent to absorbing a photon.  When the intensity field is time static, the condition that the divergence of the flux equal zero is equivalent to radiative equilibrium (Mihalas 1978).  This is the self-consistency condition of energy balance that determines the equilibrium temperature in a dust envelope.  The energy emitted by dust grains is equal to $\int \kappa_{\nu}B_{\nu}(T(\bar{x}))d\nu$ and the energy absorbed is equal to $\int \kappa_{\nu}J_{\nu}(T(\bar{x}))d\nu$, where $J_{\nu}$ is the mean intensity and $T(\bar{x})$ is the temperature at vector postion $\bar{x}$.  When the envelope is optically thick to its own reprocessed radiation, the mean intensity depends not only on incident photons from the source of radiation but also on reprocessed photons emitted by dust grains, i.e., this is a nonlocal process.  Small grains can also be subject to transient heating (Purcell 1976), with the temperature of small grains fluctuating above and below the equilibrium temperature.  We defer the treatment of stochastically heated small grains to a future paper.  We focus here on an examination of algorithms designed to enforce radiative equilibrium in Monte Carlo codes which is necessary to calculate the global radiation field within a dusty envelope.

\subsection{Iterative \& Relaxation Algorithms}

Iteration is a prevalent method in enforcing radiative equilibrium to solve for the global radiation field in finite difference codes, such as the widely used code DUSTY (Ivezic \& Elitzur 1997).  
It is far less commonly used in Monte Carlo codes, perhaps due to the preconception that it would be time-consuming.  
A straightforward and efficient algorithm to enforce radiative equilibrium in Monte Carlo codes was presented by Lucy (1999).  We demonstrate in \S 3 that this algorithm is particularly efficient for modeling three-dimensional geometries.

Lucy (1999) presented an iterative means of calculating the radiative equilibrium dust temperature, which he demonstrated produced correct results in a one-dimensional homogeneous envelope.   The key insight of this scheme is to take the Monte Carlo estimator for a volume element's mean intensity (which is proportional to the absorbed energy) to be given by the sum of path lengths of all photon packets in a given frequency range that traverse that volume element:
\begin{equation}
J_{\nu} d\nu = \frac{1}{4 \pi}\frac{\epsilon_{0}}{\Delta t} \frac{1}{V} \sum l
\label{eq:lucy_Jnu}
\end{equation}
where $\epsilon_{0}$ is the energy of the photon, $\Delta t$ is the duration of the Monte Carlo simulation, and the summation is over all photons packets in the volume V within frequencies $(\nu, \nu+d\nu)$.  It makes intuitive sense that taking the absorbed energy to be proportional to the sum of path lengths rather than the number of absorptions (see discussion below of Bjorkman \& Wood's method) will lead to a efficient estimate of the mean intensity, particularly when there are few absorptions as in large 3-D grids.  The re-emitted photon's frequency is sampled from the emissivity, i.e.,
\begin{equation}
\frac{dP_{i}}{d\nu}=\kappa_{\nu}B_{\nu}(T)
\end{equation}
The temperature calculation in this case proceeds iteratively.   One starts with a guess for the temperature; the Monte Carlo simulation is run with some number of photons for that iteration.  The temperature is then calculated on the basis of radiative equilibrium, which means that the rate of energy absorption $\dot{E}_{\rm abs}=4 \pi \int \kappa_{\nu} J_{\nu} d\nu$ is equal to the rate of emission $\dot{E}_{\rm em}=4 \pi \int \kappa_{\nu} B_{\nu} d\nu$.   Writing the mean intensity as in equation (\ref{eq:lucy_Jnu}) gives the formula for the  energy absorption rate that is given by Lucy (1999; e.g. Eqn 14 in that paper):
\begin{equation}
\dot{E}_{\rm abs}=\frac{\epsilon_{0}}{\Delta t} \frac{1}{V}\sum \kappa_{\nu} l\;.
\end{equation}
We then have the following equation to tabulate the temperature:
\begin{equation}
\sigma T_{i}^{4}=\frac{L\sum\kappa_{\nu}l}{4N_{\gamma}\kappa_{P}(T_{i})V} \;,
\end{equation}
where the quantity $\epsilon_{0}/\Delta t$ has been expressed as $L/N_{\gamma}$ where $N_{\gamma}$ is the number of photon packets and $L$ the total luminosity.

The iteration continues, with the emissivity sampled and the temperature calculated as described above for each iteration, until the 
user implements an exit criterion specified by some tolerance level for the temperature convergence.  At the point the temperatures have converged (to within some tolerance), the emitted photons will be sampled from the correct emissivity, and therefore produce the correct emergent spectrum. 
We show in the following section that monitoring the derivative of some quantity, such as the standard deviation, and exiting the iteration loop when the derivative is no longer changing (within some tolerance level) is a generically applicable scheme that holds for homogeneous and clumpy geometries.   The actual magnitude of the standard deviation does vary for various geometries and degrees of clumpiness.  Thus, what we monitor is the derivative, i.e., the change in some quantity with respect to iteration number, and end the iteration when the derivative of that quantity is no longer changing with iteration number:
\begin{equation}
V_{i+1}=\frac{\sigma_{i+1}-{\sigma}_{i}}{\sigma_{i+1}},~~~~~\rm exit:~\frac{V_{i+1}-V_{i}}{V_{i+1}} \la 0.1 \;,
\end{equation}
where $\sigma$ is the standard deviation of the temperatures of the grid cells, and $V_{i}$ is the derivative with respect to iteration number.    

The Bjorkman \& Wood method (Bjorkman \& Wood 2001; henceforth BW01) has been implemented in a number of radiative transfer codes and applications thereof (Wood et al. 2002; Whitney et al. 2003a; 2003b; Kuraskiewicz et al. 2003; Dullemond \& Dominik 2004; Dullemond \& van Bemmel 2005; Stamatellos \& Whitworth 2005; Indebetouw et al. 2006; Chakrabarti et al. 2007a).  It is a relaxation method, i.e., it is not explicitly iterative, that has been widely used to enforce radiative equilibrium in Monte Carlo codes.  We review briefly the key elements of this method.  The energy absorbed by a grid cell is taken to be given by the product of the number of absorbed photons times the energy of the photon packet.  The energy emitted by a grid cell is given as usual by the Planck function weighted by the opacity, i.e., this is not an unique feature of BW01.  A key clever element of this method is that the frequency with which photons are reemitted is sampled from a $\it{difference~function}$.  It is this procedure that allows for the temperature
and emission spectrum
to be corrected as successive photons are absorbed, without iteration.  In more detail - let $T_{i}$ denote the new temperature of a volume element after a photon has been absorbed and $\Delta T$ the temperature increase that occurs when the photon packet is absorbed.  The total energy that should now be radiated corresponds to the mean intensity evaluated at the new temperature $T_{i}$.  The energy that should be carried away is proportional to the difference in emissivities between the emissivity evaluated at the old temperature and the new temperature, i.e.,:
\begin{eqnarray*} 
\Delta j_{\nu}=j_{\nu}-j_{\nu}'=\kappa_{\nu}\left[B_{\nu}(T_{i})-B_{\nu}(T_{i}-\Delta T)\right]~~~ \\ 
\approx \kappa_{\nu}\Delta T (dB_{\nu}/dT) 
\end{eqnarray*}
where on the right hand side the difference is approximated as a derivative for small $\Delta T$.  
BW01 normalize this function and find then that the frequency of the remitted photon should be sampled from the following probability distribution function:
\begin{equation}
\frac{dP_{i}}{d\nu}=\frac{\kappa_{\nu}}{\rm K}\left(\frac{dB_{\nu}}{dT}\right)
\end{equation}
where $K=\int \kappa_{\nu}(dB_{\nu}/dT)d\nu$ and the derivative of the Planck function is 
evaluated at $T=T_{i}$.  

One can iteratively calculate the temperature taking the absorbed energy to be given by the number of photons absorbed (Misselt et al. 2001).  Although this is a more generic scheme, as it can treat stochastically heated grains, it is not as efficient as Lucy's method, which iteratively calculates the temperature taking the absorbed energy to be given not by the number of photons $\it{absorbed}$ in a grid cell, but rather by the number of photons $\it{traversing}$ a grid cell.

\section{SEDs of Homogeneous Envelopes In One \& Three-Dimensional Grids}

\subsection{SEDs of Homogeneous Envelopes in 1-D Grids}

We first discuss the Lucy and BW01 methods in the context of a spherically symmetric homogeneous density profile with a central source of luminosity in a one-dimensional grid.  In \S 3.1.1, we discuss two separate conditions for convergence, namely, the number of photons required to reach a converged temperature profile ($N_{\rm temp}$), and the number of 
photons desired for SED fidelity (which depends on the desired spectral resolution and wavelengths of interest).     
We refer to this latter convergence condition as $N_{\rm SED}$.  We require both of these conditions to be met.  We find for one-dimensional envelopes that $N_{\rm SED} \gg N_{\rm temp}$ when using either the Lucy or BW01 methods
, and thus BW01 is somewhat more efficient since it does not require iteration.  
We proceed to three-dimensional envelopes in \S 3.2.

We consider a dust envelope having an embedded central protostar with a luminosity of $101.7~L_{\odot}$, an envelope mass of $4.28~M_{\odot}$ and an outer radius of 16441.6 AU, which are parameters comparable to those of low or intermediate mass protostars.  The density varies as $r^{-1.5}$ within the envelope.  We have used the Weingartner \& Draine (2001) dust opacity curve, with scattering efficiencies set to zero for simplicity, specifically to highlight the effects of clumpiness and the diffuse phase on the near and mid-IR part of the spectrum in \S 3.3 (note however we do consider scattering in our galaxy SED calculations in \S 4).  As shown by Chakrabarti \& McKee (2005; henceforth CM05), for a given density profile and dust opacity curve, the luminosity, mass and size determine the emergent (far-IR) SED.  The stellar temperature and radius are $5\times 10^{3} \rm K$ and $13.47~R_{\odot}$ respectively.  We have used a spherical polar grid, with logarithmic spacing in the radial direction.  The total number of cells is 400.  

\subsubsection{Number of Photons Required for Temperature ($N_{\rm temp}$) and SED Convergence ($N_{\rm SED}$) in 1-D Grids}

\begin{figure}[!ht] \begin{center}
\centerline{\psfig{file=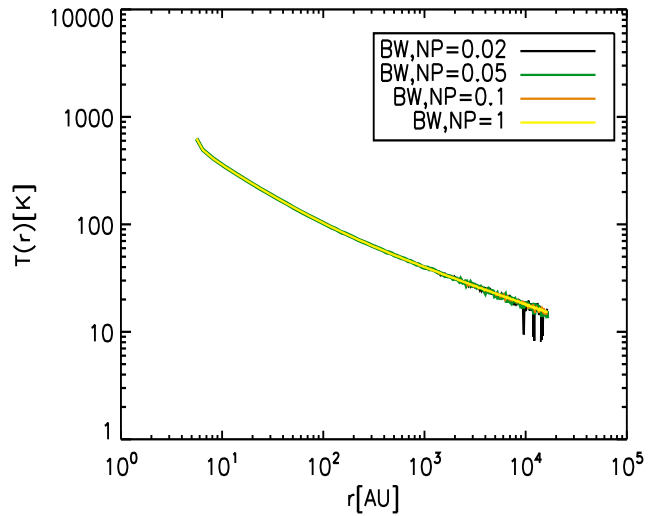,height=3.in,width=3.5in}}
\end{center}
\caption{Quantifying $N_{\rm temp}$ for homogeneous envelopes in 1-d grids using BW01's temperature algorithm - the number of photons (NP) is given in units of $10^{6}$.  For this 1-d grid, there are 400 grid cells, and 50,000 photons (i.e., $N_{\rm temp}=125 N_{\rm grid}$) are sufficient to reach a converged temperature profile throughout the envelope, and 20,000 photons (i.e., $N_{\rm temp}=50 N_{\rm grid}$) sufficient to reach a converged temperature profile everywhere except the very outer regions.}
\end{figure}

We show in Figure 1 the calculated radiative equilibrium temperature profile in this 1-d grid
using the BW01 method.  The photon number is varied from $20,000-10^{6}$ photons.
As is clear, 50,000 photons, which corresponds to $125 N_{\rm grid}$, is sufficient to reach a converged temperature profile everywhere in the envelope.  Even 20,000 photons, which corresponds to $50 N_{\rm grid}$, is sufficient to reach a converged temperature everywhere in the envelope except the very outer regions.  These calculations take 23s and 19s of CPU time on a 2 GHz AMD Opteron processor respectively. In summary, $N_{\rm temp} \sim 100 N_{\rm grid}$ produces a converged temperature profile throughout the dust envelope.

Figure 2 depicts the emergent SED from this dust envelope using the BW01 method with $1,~5,~10\times 10^{6}$ photons.  The SED is converged close to peak wavelengths even with $1 \times 10^{6}$ photons which requires only 136.87s of CPU time on a 2 GHz AMD Opteron processor.  However, the longwave SED is somewhat noisy for this photon number.  With 5 million photons (or higher), even the longwave SED is smooth, for which the CPU time is 550s, i.e., the CPU time scales nearly linearly with photon number for the BW01 method. Therefore, the number of photons required to reach a noise-free long wave SED, which is about a million, is about two thousand times the number of grid cells.  This is considerably greater than the number of photons to reach a converged temperature profile, i.e, $N_{\rm SED} \gg N_{\rm temp}$.  

\begin{figure}[!ht] \begin{center}
\centerline{\psfig{file=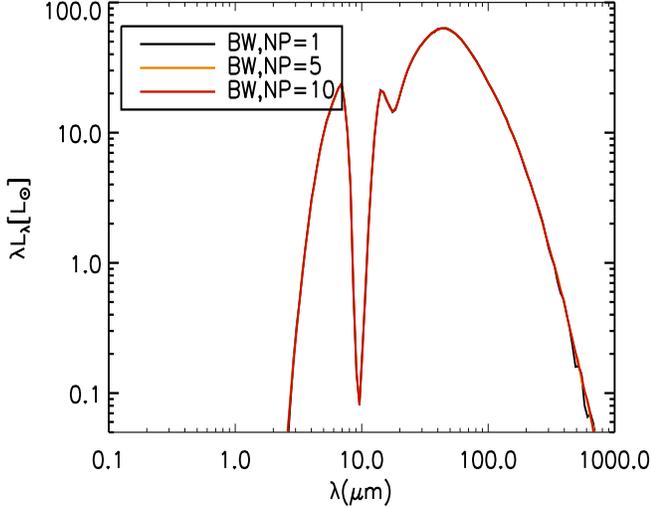,height=3.in,width=3.5in}}
\end{center}
\caption{Emergent SED from one-dimensional envelope for a spherically symmetric density profile using Bjorkman \& Wood method with 1,5,and 10 million photons (the number of photons (NP) is given in units of $10^{6}$).  $N_{\rm SED}$, i.e., the number of photons required to reach a converged SED, is much higher than $N_{\rm temp}$.  1 million is nearly sufficient and 5 million sufficient to reach a converged long wave SED.}
\end{figure}

Exiting the iteration loop when using the Lucy method with various exit criteria is discussed in Appendix I.   We adopt the $V4$ exit criterion, which exits with respect to a standard deviation, taking the average with respect to the difference of temperatures in all grid cells in the $i+1$ iteration and the i'th iteration.   We adopt $V4$ as our fiducial iteration exit criterion since as a standard deviation, it incorporates information both about the average of and the relative spread in the temperatures per iteration.  It is also efficient, requiring 772.10 of CPU time on a 2 GHz AMD Opteron processor.  

We quantify $N_{\rm temp}$, i.e., the number of photons to reach a converged temperature profile, when using the fiducial iterative $V4$ Lucy method, in Figure 3.  A thousand photons per iteration is sufficient to reach a converged temperature profile everywhere in the envelope, which corresponds to $N_{\rm temp}=2 N_{\rm grid}$ per iteration and $N_{\rm temp,total}=12 N_{\rm grid}$.  Calculations done with fewer photons per iteration do not converge within our tolerance limit.  Iterating with 1000 photons took less than 10 s of CPU time on a 2 GHz AMD Opteron processor.  

\begin{figure}[!ht] \begin{center}
\centerline{\psfig{file=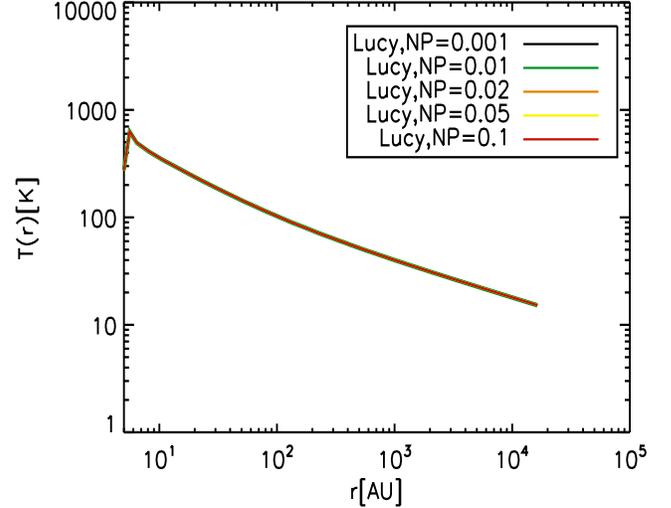,height=3.in,width=3.5in}}
\end{center}
\caption{Quantifying $N_{\rm temp}$ for homogeneous envelopes in 1-d grids using the $V4$ fiducial iterative algorithm - the number of photons (NP) is given in units of $10^{6}$.  For this 1-d grid, there are 400 grid cells, and 1000 photons per iteration (i.e., a total of 5000 photons and $N_{\rm temp,total}=12 N_{\rm grid}$ or $N_{\rm temp}=2 N_{\rm grid}$ per iteration)  are sufficient to reach a converged temperature profile throughout the envelope.}
\end{figure}

\begin{figure}[!ht] \begin{center}
\centerline{\psfig{file=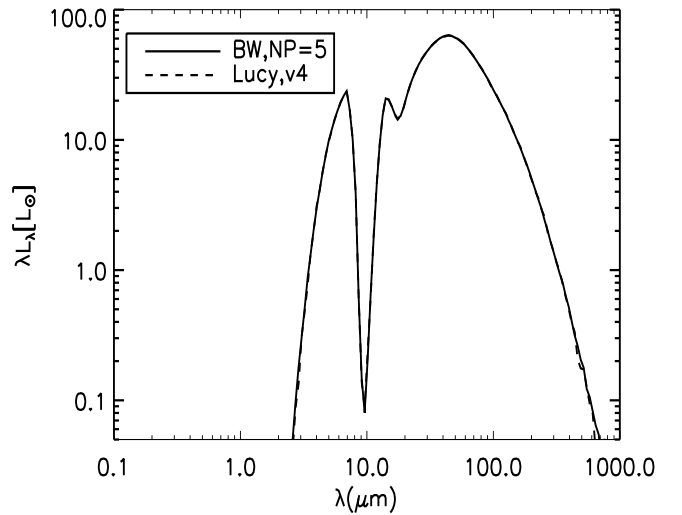,height=3.in,width=3.5in}}
\end{center}
\caption{Emergent SED from one-dimensional envelope for a spherically symmetric density profile using the iterative Lucy method and exiting by reference to the standard deviation (v4) (dashed line), compared to the Bjorkman \& Wood method (solid line) with 5 million photons, as shown above in Figure 2.}
\end{figure}

Figure 4 shows that the Lucy method (exiting with respect to the standard deviation, $V4$, and using one million photons per iteration which took 777s of CPU time) and the BW01 method (run with $5 \times 10^{6}$ photons) yield identical emergent SEDs for this one-dimensional, homogeneous test case.  Again, $N_{\rm SED} \gg N_{\rm temp}$ when using the Lucy method also.  In summary, both of these methods are nearly comparable in efficiency for use in computing the emergent SEDs of one-dimensional envelopes, with the BW01 method being somewhat faster since it does not require iteration.  (Temperature convergence is achieved more quickly with Lucy's method in 1-D grids.   However, unlike the SED, the temperature is not a quantity that can be directly compared to observed data).

SEDs of two-dimensional homogeneous envelopes have been studied extensively by Whitney et al. (2003a;2003b).  We have learned from experience using our 2-D models of Young
Stellar Objects (Whitney et al. 2003a,b) that $N_{\rm temp}$
and $N_{\rm SED}$ are approximately comparable using the BW01 method.
The standard setup for these models has 400x199=79,600 grid cells, i.e.,
400 radial cells and 199 $\theta$ cells.
$N_{\rm SED}$ is about 1 million when we use the option to produce a high
signal-to-noise
SED in a single direction.  For producing SEDs at ten viewing angles,
$N_{\rm SED} \ga 10 \times 10^{6}$.
We find that $N_{\rm temp}$ is about 4 million, or $50 N_{\rm grid}$); using lower
values gives the characteristic lower long-wave
SED because the cell temperatures have not fully relaxed and are too
low.
Thus in 2-D models, the BW01 and Lucy method have comparable
efficiencies for computing high-resolution SEDs.
However, the Lucy method has a much lower $N_{\rm temp}$, so in cases where the requirements
on $N_{\rm SED}$ are lower (i.e., broadband fluxes near the peak of the SED), 
this method will be more efficient.

\subsection{SEDs of Homogeneous Envelopes in 3-D Grids}

We now examine the results for a spherically symmetric homogeneous density profile in a three-dimensional grid;
that is, we are computing 1-D models in a 3-D grid to examine results from the two temperature algorithms as the number of grid cells is increased.  
The parameters are the same as before, but we have now used 200 radial cells, 59 grid cells in $\theta$ and 118 in $\phi$.  This gives a total of $1.39\times 10^{6}$ grid cells.  In Appendix II we discuss some details of implementing the iterative Lucy method, including how the resultant SED depends on the tolerance and photon number per iteration.

\subsubsection{Number of Photons Required for Temperature ($N_{\rm temp}$) and SED Convergence ($N_{\rm SED}$) in 3-D Grids}

\begin{figure}[!hb] \begin{center}
\centerline{\psfig{file=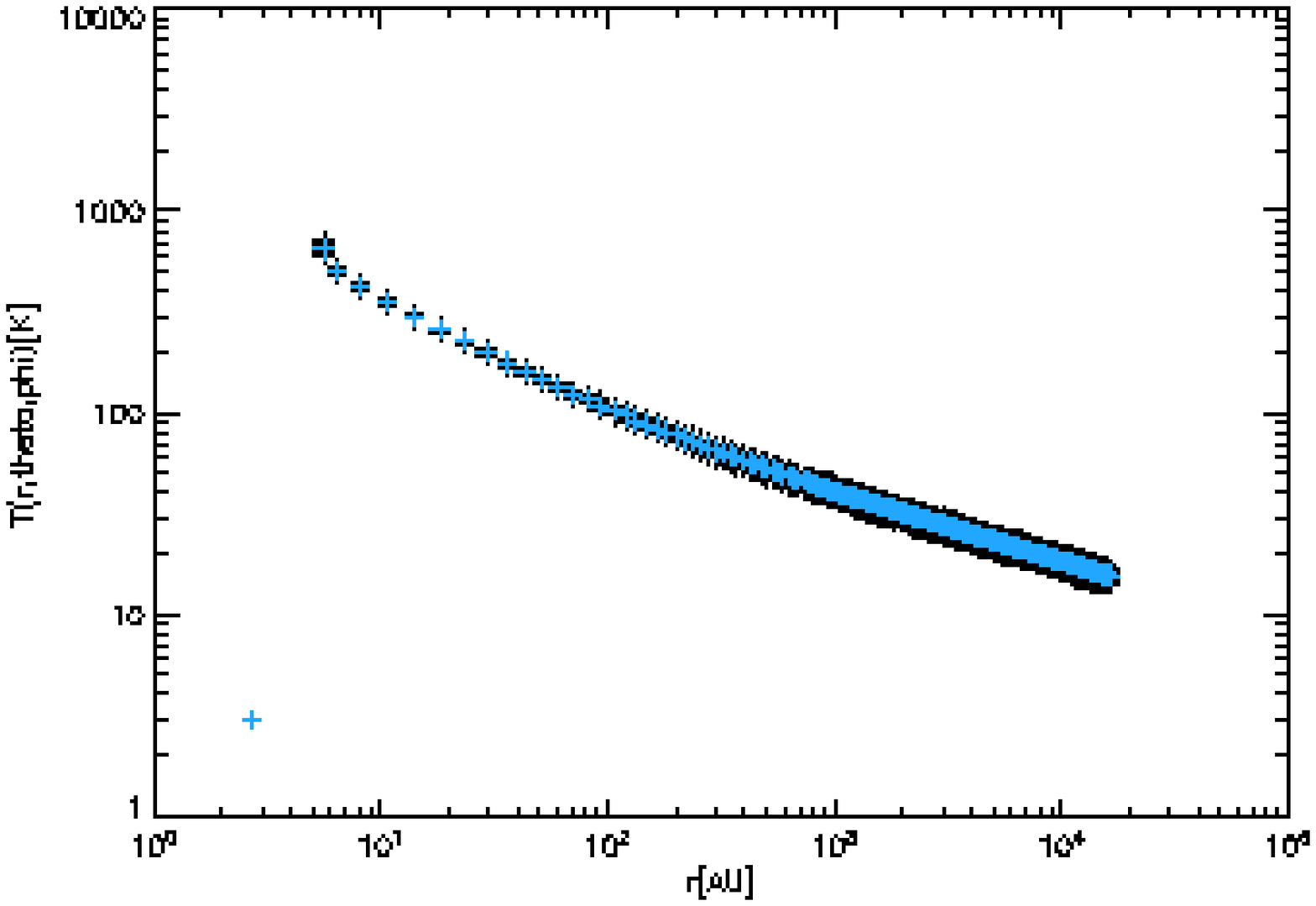,height=2.in,width=2.in}
\psfig{file=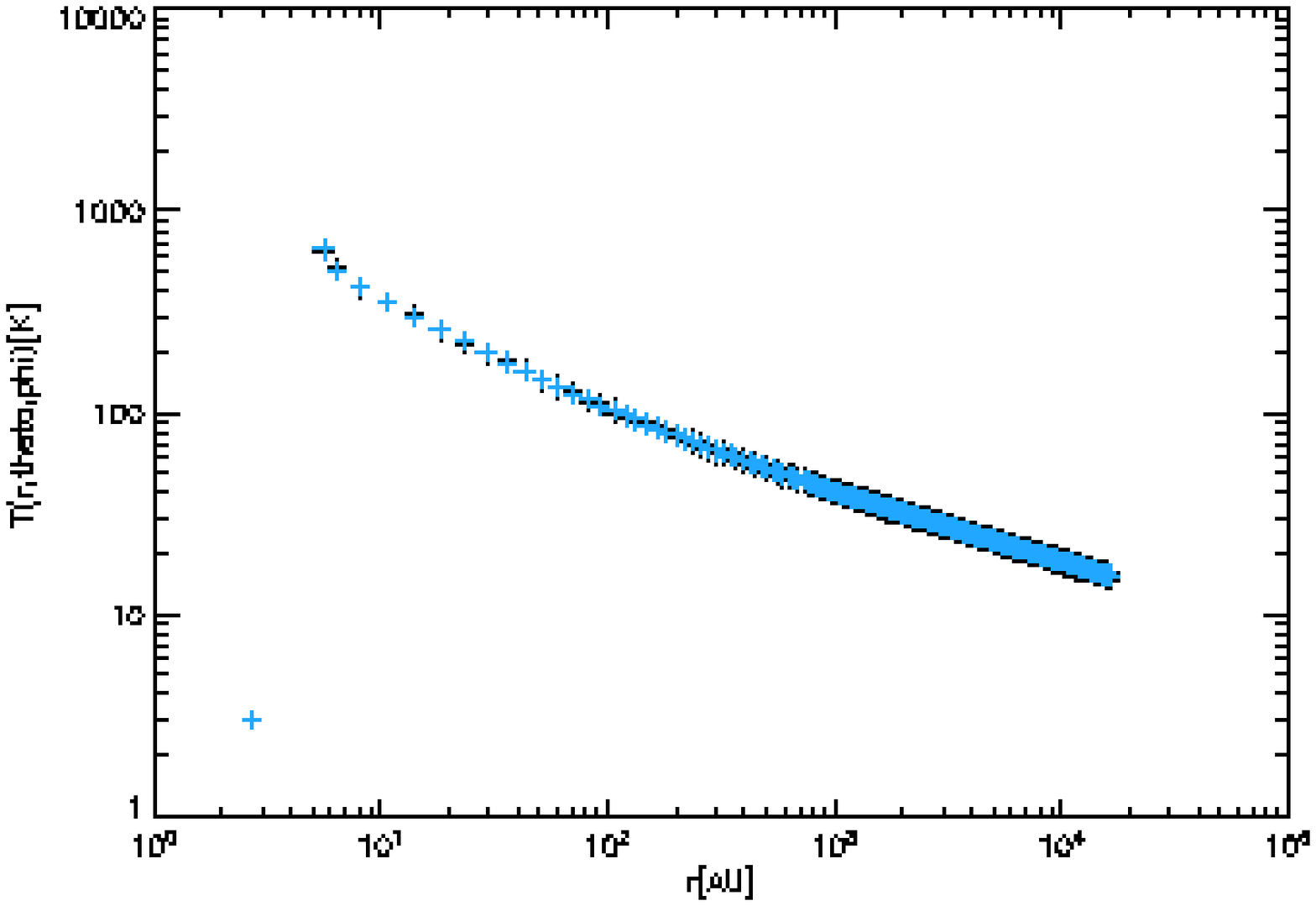,height=2.in,width=2.in}}
\end{center}
\caption{(a) Resultant temperature profile using the fiducial iterative scheme with 1 million photons per iteration for the SED shown in Figure 11.  The colored line denotes the average temperature. (b) Resultant temperature profile using the fiducial iterative scheme with 6 million photons per iteration.}
\end{figure}

\begin{figure}[!hb] \begin{center}
\centerline{\psfig{file=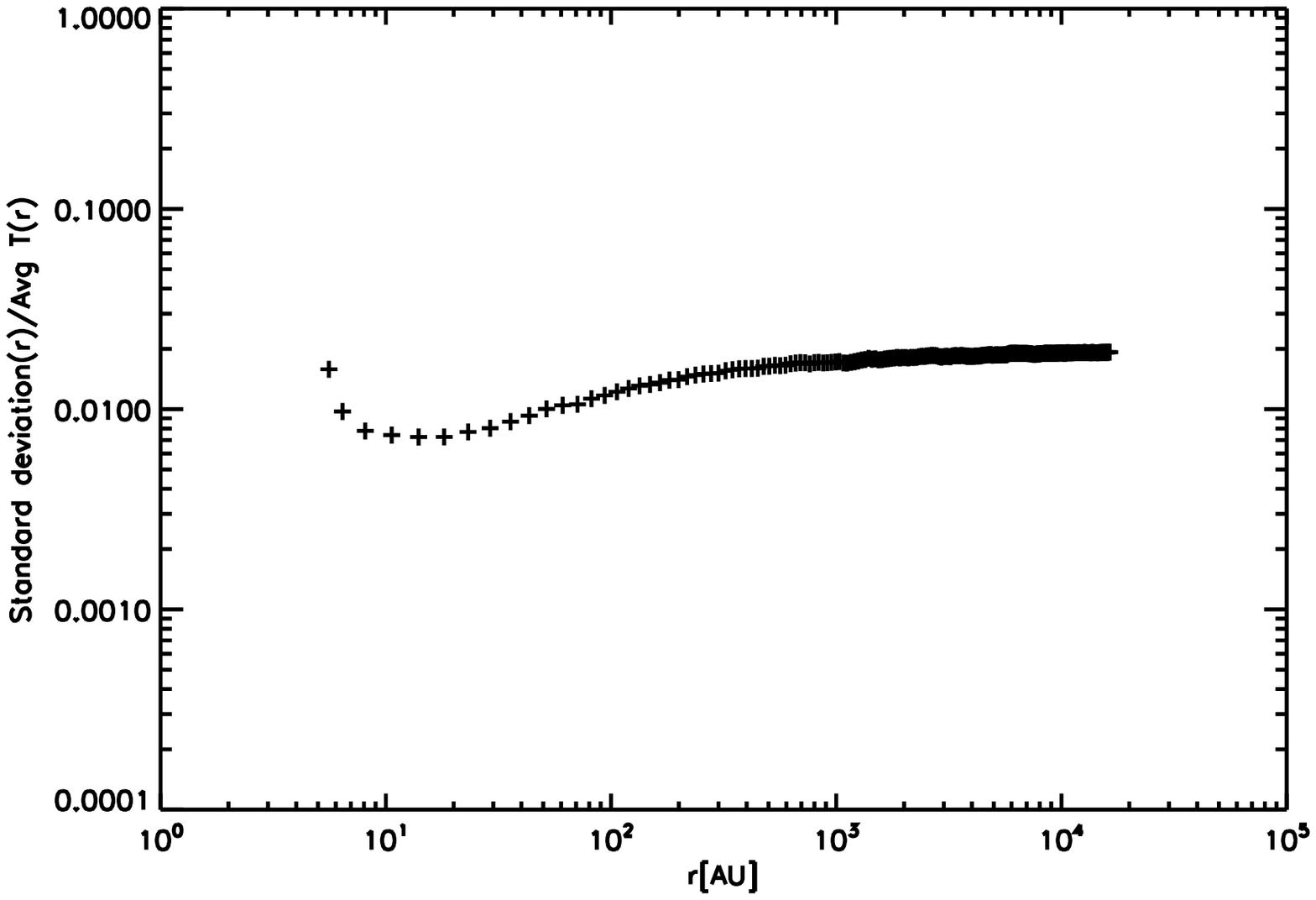,height=2.in,width=2.in}
\psfig{file=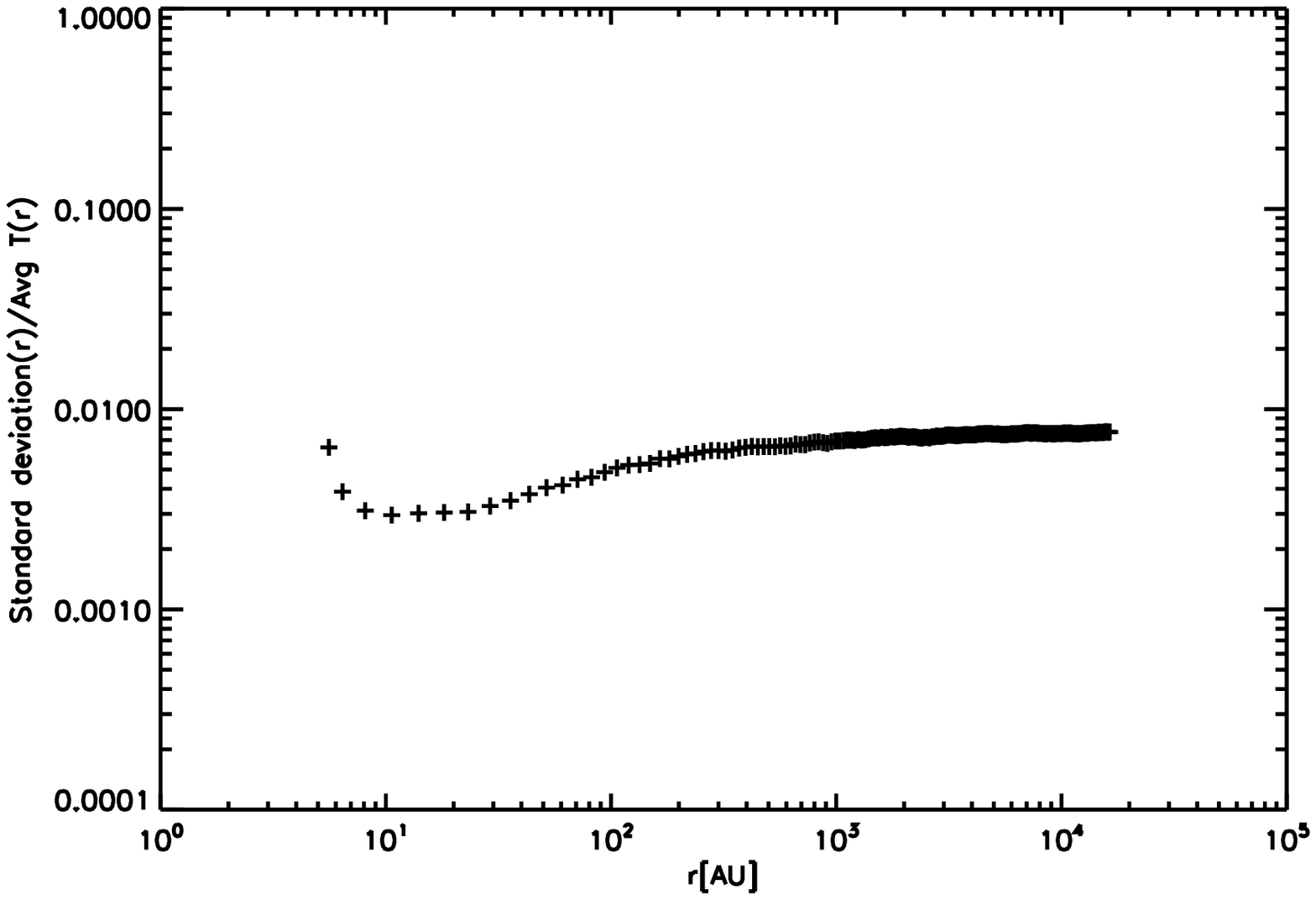,height=2.in,width=2.in}}
\end{center}
\caption{(a) Ratio of the standard deviation of the temperature with respect to the average temperaturefor the case shown above, using the fiducial iterative scheme with 1 million photons per iteration.  In this case, the maximum of this ratio is equal to 0.019.  (b) Ratio of the standard deviation of the temperature with respect to the average temperature for the case shown above, using the fiducial iterative scheme with 6 million photons per iteration.  Increasing the photon number lowers the ratio below 1\%.}
\end{figure}

To get a quantitative handle on the number of photons required to reach a converged temperature profile, we require that
at each radius, the ratio of the standard deviation of the temperatures 
in each $\phi$ and $\theta$ grid cell to the average temperature be less than 2 \%.  
The temperature profile of homogeneous envelopes that are modeled in a 3-D grid should have no dependence on angle.  
Thus, the variation reflects the error in the temperature calculation.
The calculated radiative equilibrium dust temperature for our model 
using the fiducial Lucy method 
is depicted in Figure 5 a and 5 b when 1 million and 6 million photons are used respectively per iteration.  As is clear, increasing the photon number decreases the noise in the temperature calculation.  The ratio of the standard deviation with respect to the average temperature (which is shown in color in Figures 5) is shown in Figure 6 a and 6 b.  Using 1 million photons per iteration is sufficient to reach a converged temperature profile at the level of $< 2$\% accuracy for $N_{\rm grid}=1.39 \times 10^{6}$.  In general, we find that $N_{\rm temp,per~iteration}=0.72 \times N_{\rm grid}$ is sufficient to reach a converged temperature at the level of 2 \% using the fiducial Lucy method.  A total of four iterations are typically required to reach this level of accuracy, i.e., a total number of photons $N_{\rm temp,total}=2.87 \times N_{\rm grid}$ is required to reach a 2 \% level of accuracy in the temperature.  

We show in Figures 7 and 8 the analogous quantities when the BW01 temperature algorithm is used.  The temperatures are considerably noisier when this relaxation method is used in three-dimensional grids with a large number of grid cells (where large is $\ga 10^{6}$).  Here, the total number of photons that are been used in panels (a) and (b) in these two figures is $80 \times 10^{6}$ and $320 \times 10^{6}$ respectively.  As is clear, for the BW01 method, even $N_{\rm temp, total} \sim 320 N_{\rm grid}$ is not sufficient to reach 2 \% accuracy in the temperature profile throughout the dust envelope.  There is also a clear difference in the ratio of the radial variation of the standard deviation of the temperatures to the average temperature ($\sigma(r)/T_{\rm avg}(r)$) when the BW01 and Lucy algorithms are used.  When the BW01 method is used, $\sigma(r)/T_{\rm avg}(r)$ is very accurate in the inner regions of the dust envelope, i.e.,$\sigma(r)/T_{\rm avg}(r) < 2$ \% even when $N_{\rm temp,total}=60 N_{\rm grid}$.  However, $\sigma(r)/T_{\rm avg}(r)$ increases with radius for the BW01 method and is of order $\sim 10$\% in the outer regions of the dust envelope even with a large number of photons.  We explain below why this is the case.  Qualitatively, we can see from Figures 8 that the accuracy in the temperature when the BW01 method is used is quite good close to the Rosseland photosphere, which in this case is at $\sim 100~\rm AU$ (see e.g. CM05).  Thus, the energy density (and temperature) close to the Rosseland photosphere, which is where most the energy comes from in the dust envelope, is calculated correctly.  However, the BW01 temperature calculation becomes increasingly more inaccurate at radii far from the Rosseland photosphere, which contribute primarily to the long wavelength region of the SED.  

\begin{figure}[!ht] \begin{center}
\centerline{\psfig{file=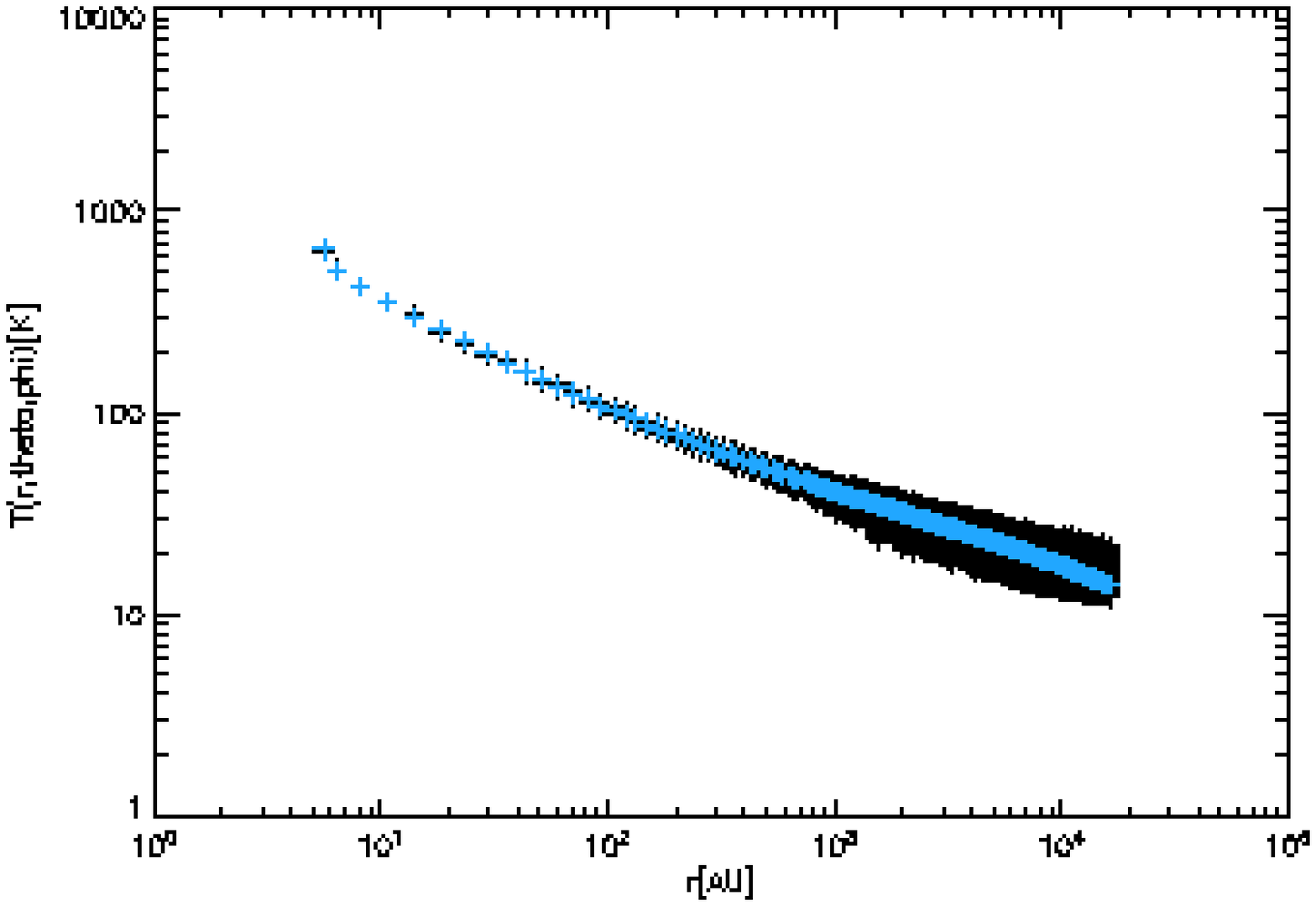,height=2.in,width=2.in}
\psfig{file=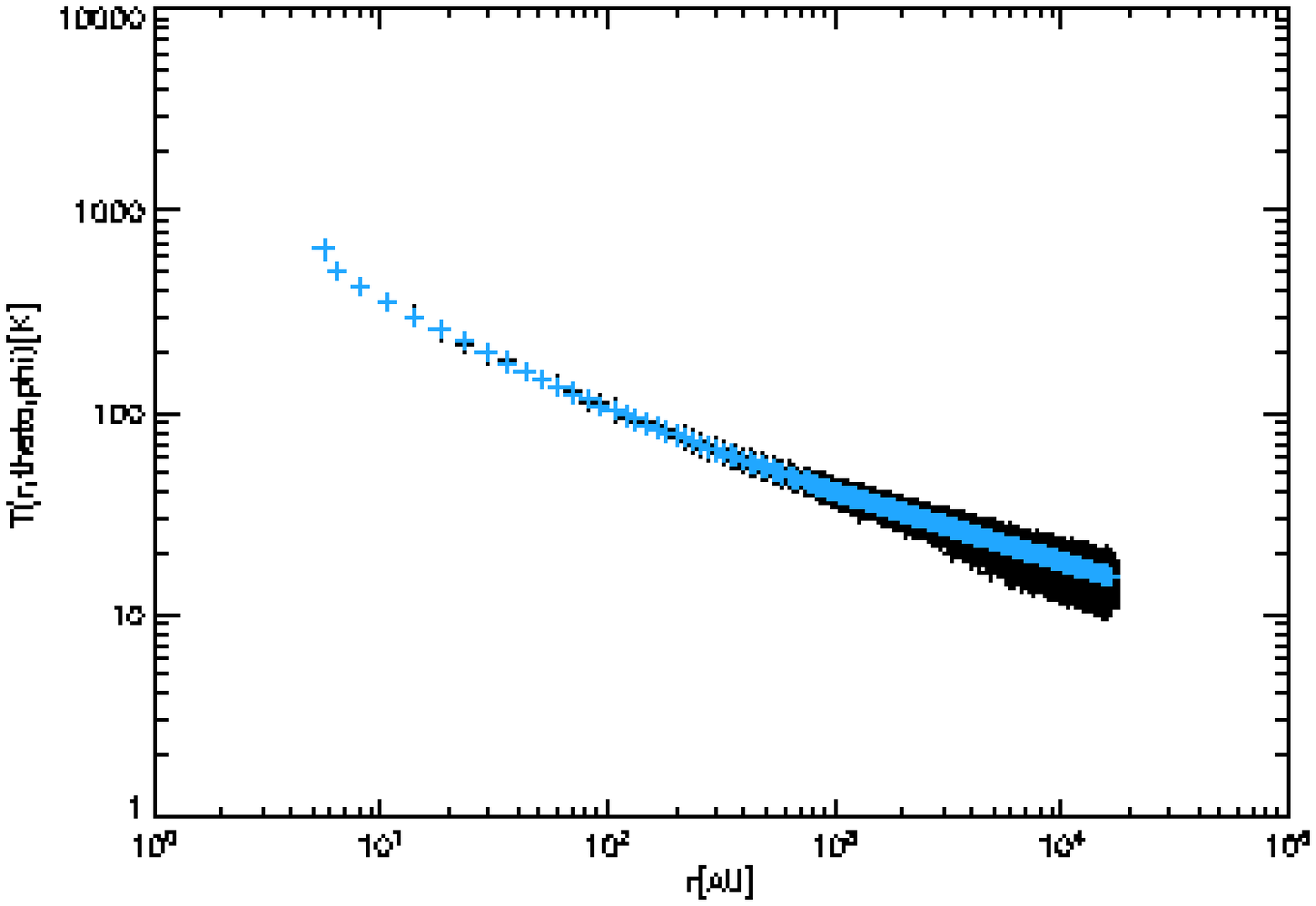,height=2.in,width=2.in}}
\end{center}
\caption{(a) Resultant temperature profile using BW01's scheme with 80 million photons.  The colored line denotes the average temperature. (b) Resultant temperature profile using BW01's temperature algorithm with 320 million photons per iteration.}
\end{figure}

\begin{figure}[!hb] \begin{center}
\centerline{\psfig{file=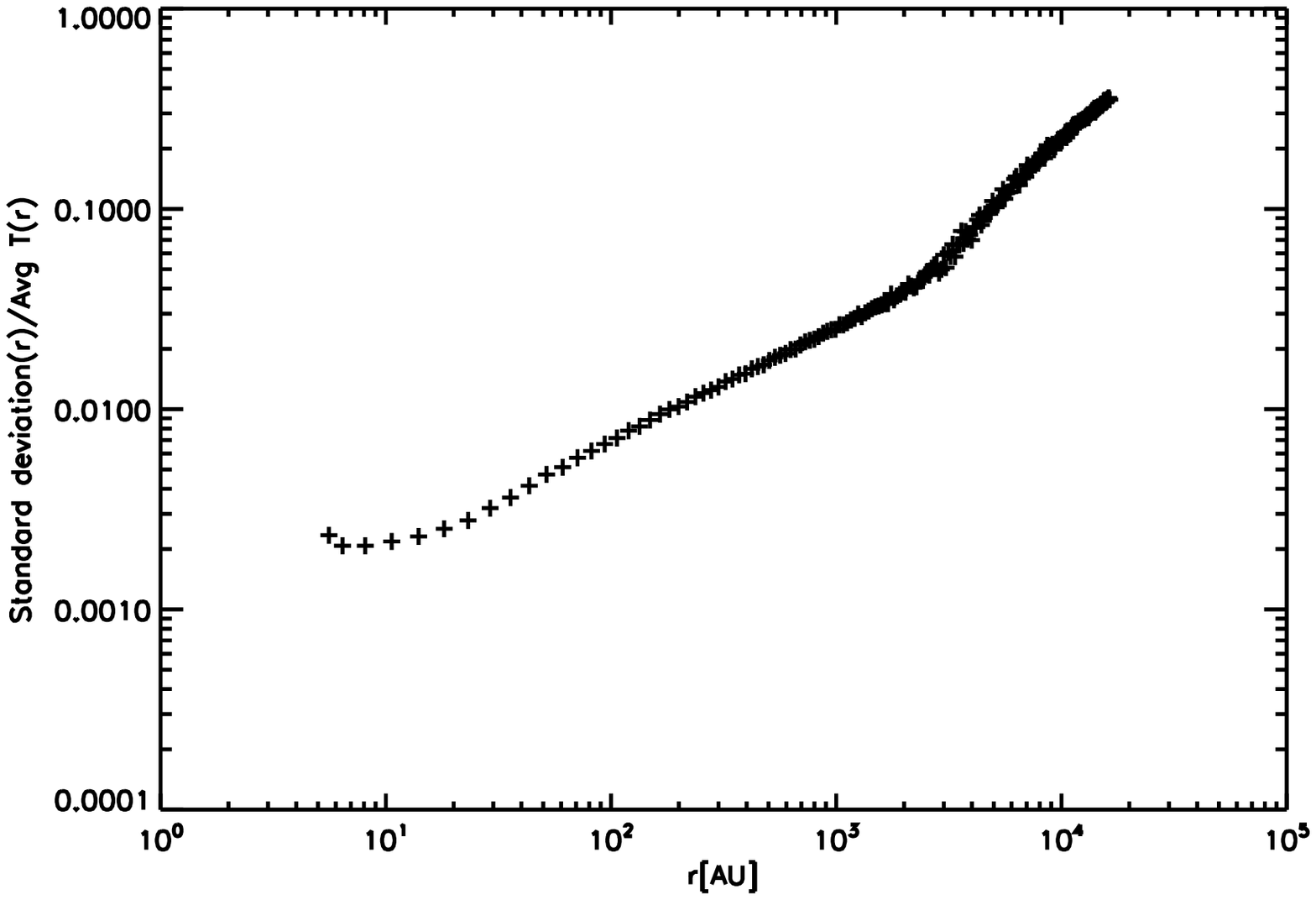,height=2.in,width=2.in}
\psfig{file=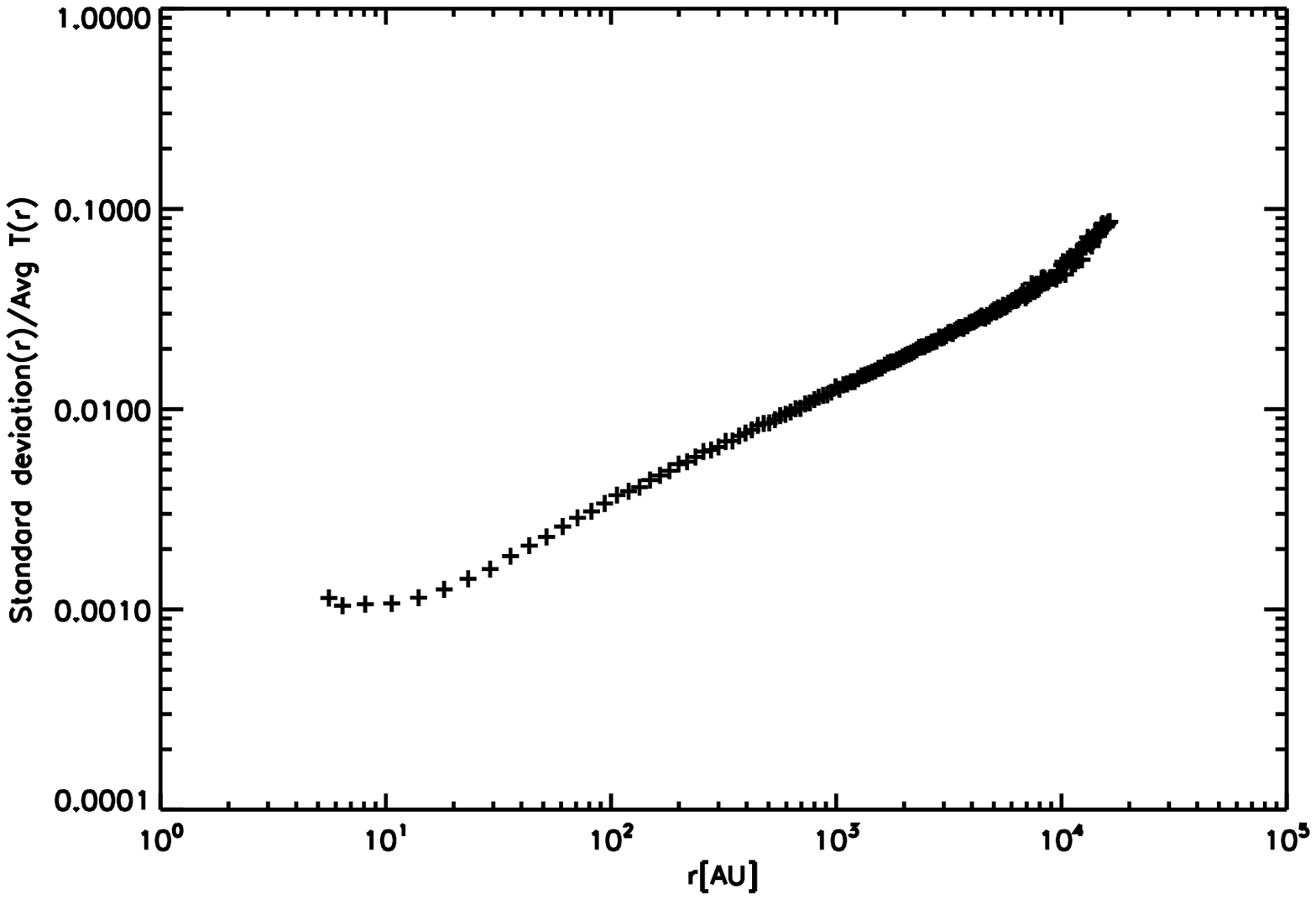,height=2.in,width=2.in}}
\end{center}
\caption{(a) Ratio of the standard deviation of the temperature with respect to the average temperature for the case using the BW01 method with 80 million photons. (b) Ratio of the standard deviation of the temperature with respect to the average temperature using the BW01 method with 320 million photons.  Note the variation with radius of this quantity - see discussion in \S 3.2.1.}
\end{figure}

Our result that $N_{\rm temp} \sim 100 N_{\rm grid}$ for $\sim$ 2\% accuracy in the temperature (close to the Rosseland photosphere) when the BW01 method is used can be understood in a simple way.  If $\sim 100$ photons are absorbed in a grid cell, photon noise ($\propto \sqrt N$) would produce an energy density that is accurate to 10 \% and a temperature (which is proportional to the 1/4th power of the energy density) that is accurate to $\sim$ 1.8 \%.  Lucy's method is faster by at least a factor of 10 (if we compare the ratio of the standard deviation to the average temperature close to the Rosseland photosphere, and faster still if we require temperature convergence in the outer parts of the envelope).  
Many more photons $\it{pass~through}$ a grid cell than are $\it{absorbed}$.
As such, Lucy's insight of estimating the mean intensity from photons traversing grid cells rather than the number of photons absorbed speeds up the computation of the temperature considerably.

The reason a much larger number of photons needs to be used for the three-dimensional grid compared to the one-dimensional grid when using the BW01 method is that in the BW01 method, the energy absorbed by a grid cell and the resultant temperature immediately following the absorption of the photon is proportional to the number of photons absorbed.  By increasing the number of grid cells, a larger total number of photons is needed to course through the envelope so that each grid cell can on average have several absorptions.  If this is not done, i.e., if the number of photons is not increased in proportion to the number of grid cells, this leads to an artificially low emissivity, particularly so in the long wavlengths.  The reason that convergence is reached more slowly for the long wavelengths is that the absorbed photon is re-emitted at a frequency that is sampled from the difference in emissivities (see the discussion earlier in \S 2).  This leads to a frequency dependence for the probability distribution function (PDF) that scales with frequency as $\nu^{3}$ (modulo $\kappa_{\nu}$ in the low frequency limit, i.e., $h\nu \ll kT$) when the PDF is sampled from the $\it{difference}$ of the emissivity.  Thus, higher frequencies converge more quickly than low frequencies.  In contrast, the Lucy method samples the emissivity, which in the low frequency limit, is proportional to $\nu^{2}$, rendering the sampling in frequency more equal by one power of frequency.  This is why Figures 6 show a roughly constant variation in $\sigma(r)/T_{\rm avg}(r)$ when the Lucy method is used while Figures 8 show the variation in $\sigma(r)/T_{\rm avg}(r)$ increasing with radius when the BW01 method is used.  

\begin{figure}[!ht] \begin{center}
\centerline{\psfig{file=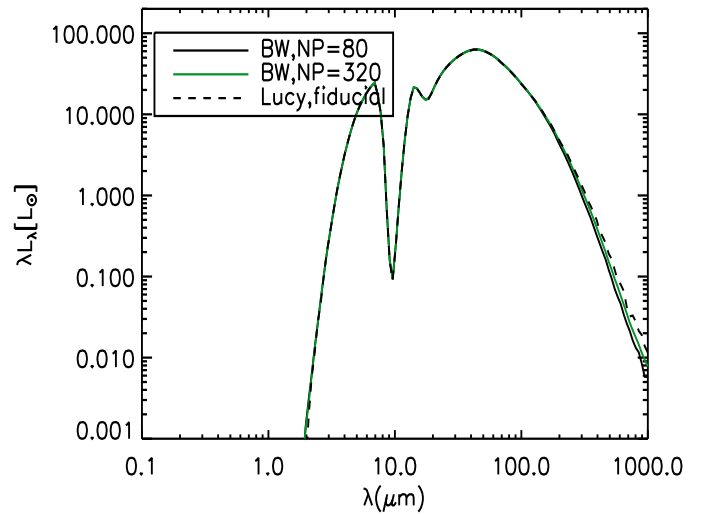,height=3.in,width=3.5in}}
\end{center}
\caption{Emergent SED averaged over 200 viewing angles from three-dimensional envelope using the Bjorkman \& Wood method with 80 and 320 million photons, compared to the fiducial iterative scheme.  The fiducial iterative scheme is 35 times faster than Bjorkman \& Wood's method in this case.}
\end{figure}

A separate convergence criterion that a user may be interested in the level of noise in the emergent SED ($N_{\rm SED}$).  In our calculations, we use $N_{\rm temp}=0.7 N_{\rm grid}$ per iteration which is sufficient to guarantee (slightly better than) 2 \% accuracy in the temperature.  As discussed previously, we have used the $V4$ iterative scheme.  When our convergence condition is satisfied, we perform one last iteration with a larger number of photons such that the level of noise in the emergent SED in the long wavelengths is acceptable.  How large the number of photons should be in this last iteration depends on what the user is primarily interested in.  If the user is interested in SEDs averaged over inclinations, then using 5 times as many photons in the last iteration is sufficient (Figure 9).  Figure 9 compares the emergent SED averaged over 200 viewing angles using the fiducial Lucy method (run with with 1 million photons per iteration and one final time with 5 million photons after the temperatures have converged) with that found using BW01's method with 80 and 320 million photons respectively.  As is clear, using 320 million photons with BW01's method yields nearly converged results.   The BW01 method took 59,618s of CPU time for this 3-D envelope, i.e., it is about 35 times slower than our fiducial iterative scheme, when comparing SEDs averaged over viewing lines of sight.  Even at this level, the SED generated by BW01's method is still slightly lower at the longer wavelengths (by $\sim 1.5$), and a further increase of a factor of 1.5 in the photon number would be required to get full agreement.  Therefore, for strict agreement out to millimeter wavelengths, BW01's method would be $\sim 50$ times slower than Lucy's method for this 3-D grid. This applies if one is interested in a SED averaged over inclinations - for SEDs at 200 viewing angles at comparable accuracy, a correspondingly larger number of photons need to used even when using the Lucy method.

In Figures 10, we show the emergent SED over 200 viewing angles using the fiducial Lucy method with 1 million photons per iteration, varying the photon number in the last iteration to improve SED fidelity.  If the user is interested in broadband fluxes within a factor of $\sim 3$ of the peak of the SED, then using eight times as many photons in the last iteration is more than sufficient (Figure 10a).  If the user is interested in a low level of noise in the sub-millimeter or millimeter wavelengths (where the energy is lower and thus fewer photons are emitted there), then a larger number of photons in the final iteration is required.  Figure 10 b shows the emergent SED over 200 viewing angles when 24 times as many photons is used in the last iteration.  If the user desires the same level of SED fidelity over all 200 viewing angles (even out to millimeter wavelengths) as in the inclination averaged SED shown in Figure 9, the photon number in the last iteration will have to increased in proportion to the number of viewing angles.   However, the millimeter wavelengths are not subject to the effects of optical depth, and as such, it is acceptable to average over viewing angles - leading to a much lower requirement for $N_{\rm SED}$.  (If comparable accuracy over hundreds of viewing angles were to be required out to millimeter wavelengths, then BW01's method could begin to approach the efficiency of Lucy's.  See however Lucy 1999b for an approach where emergent intensities are derived from formal integration that accelerate convergence over many viewing angles.  An alternate approach is to calculate a high signal to noise SED over one viewing angle in the final iteration).  In general, for most cases of interest in three-dimensional grids, the Lucy method is considerably more efficient.

\begin{figure}[!ht] \begin{center}
\centerline{\psfig{file=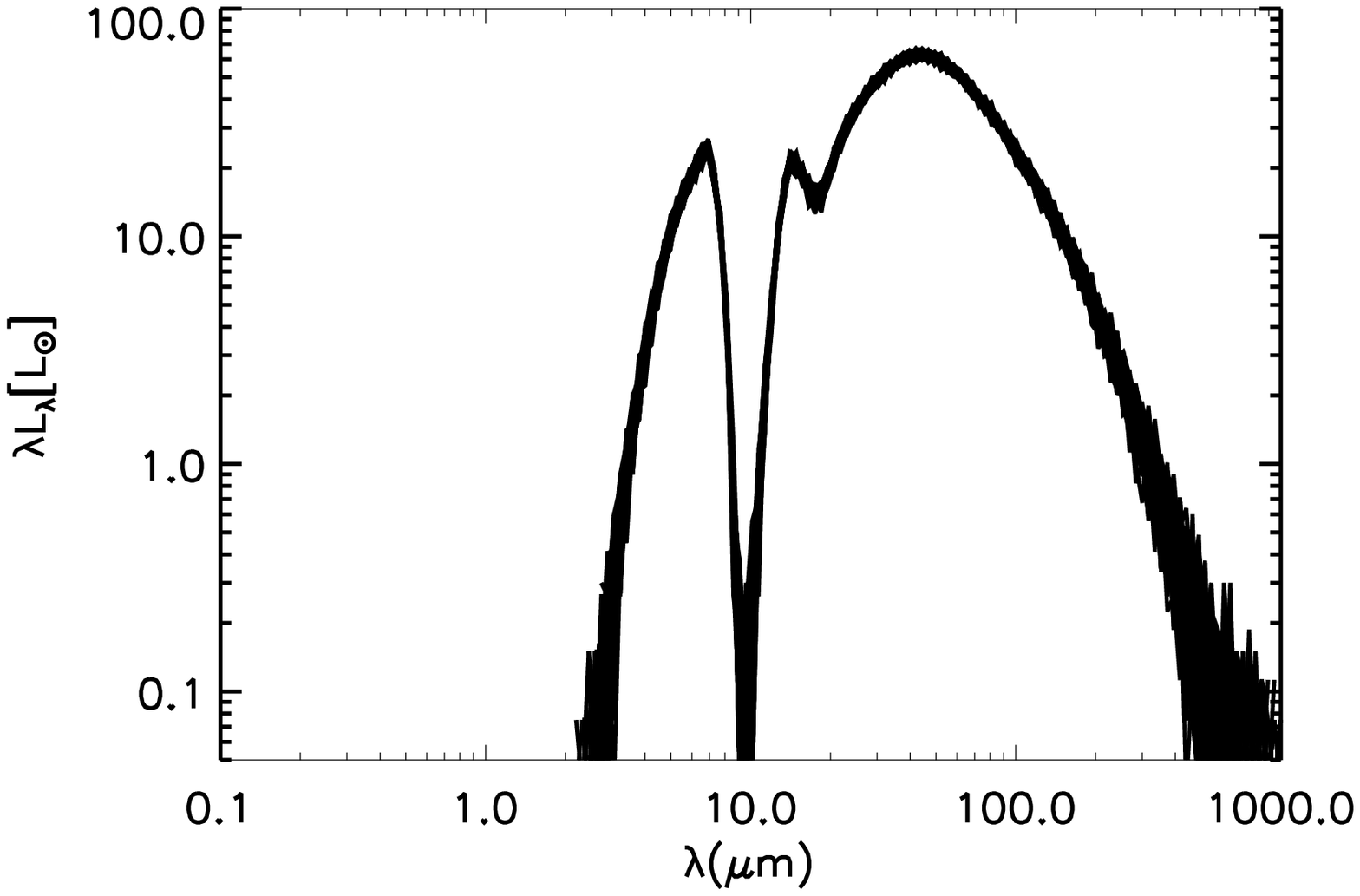,height=2.in,width=2.in}
\psfig{file=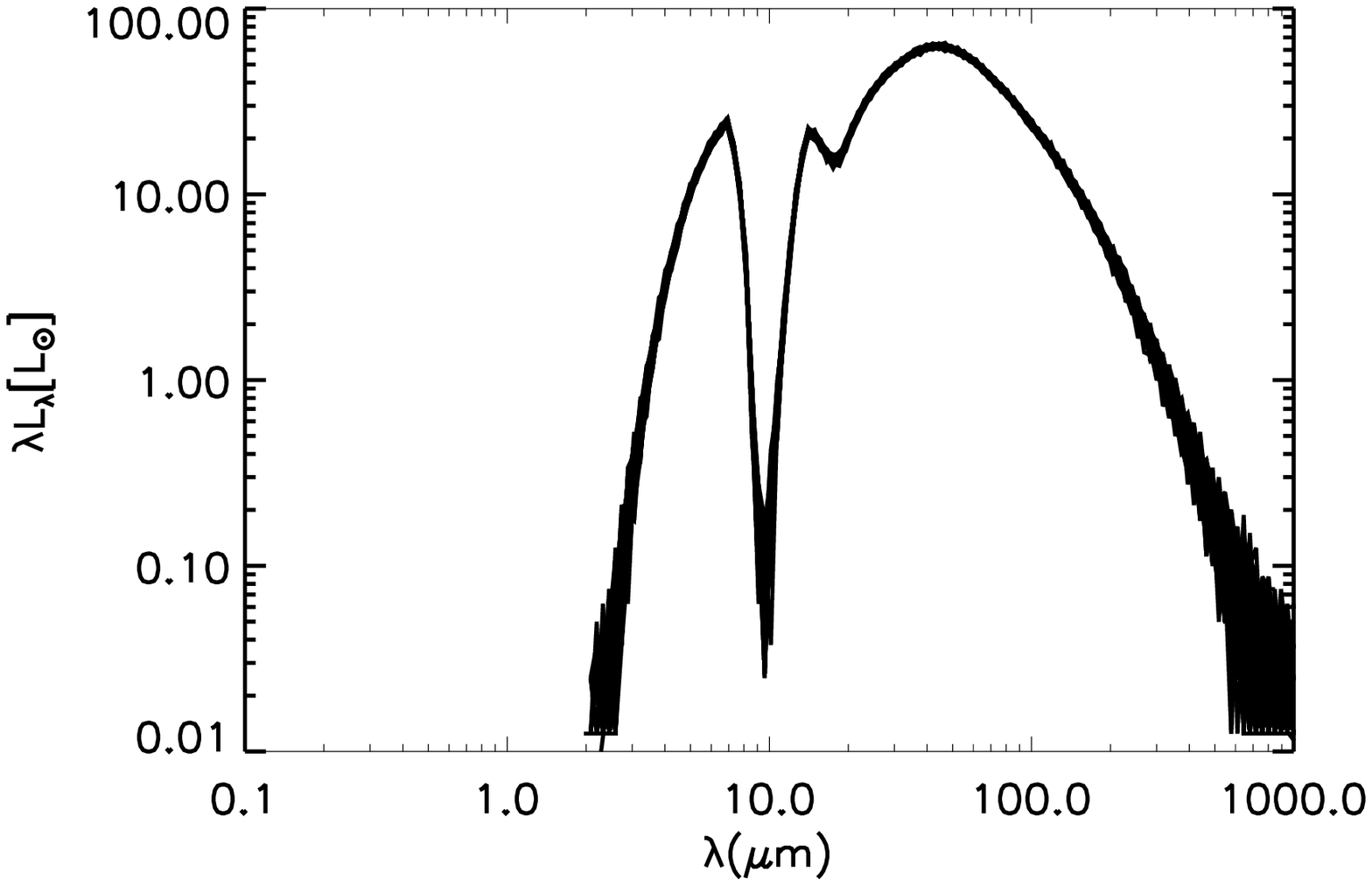,height=2.in,width=2.in}}
\end{center}
\caption{(a) Emergent SED computed using our fiducial iterative scheme, shown here over 200 viewing angles when the number of photons used in the last iteration is increased to 8 million, with one million photons used per iteration until the temperature is converged at $\sim 2$\% (b) same as (a), but with the number of photons in the last iteration increased to 24 million, the long wavelength noise in the emergent SED is lowered.}
\end{figure}

A important point of difference between BW01's and Lucy's algorithms that should be underscored is that photons traversing a low (or zero) density region contribute to the mean intensity in Lucy's method, but not to number of absorptions (which is the relevant quantity that determines the heating rate in BW01's method).  This is a reason why taking the sum of path lengths is a better estimator, i.e., a more efficient one, than the number of absorptions.  However, in the limit of large number of photons absorbed, there will eventually be enough absorptions so that the rate of photons absorbed by tracking the path-lengths of the photons traversing a grid cell is comparable to the number of absorptions.

Baes et al. (2005) have discussed the BW01 algorithm as being a special case of frequency distribution adjustment algorithms (FDA).  They show that the adjustment PDF, which depends on the temperature change, $\Delta T/T$, upon absorbing a photon packet, and the number of packets previously absorbed, $k$, is not positive for all $\Delta T/T,k$ values.  In particular, larg(er) values of $\Delta T/T$ cause the adjustment PDF to become negative at which point no further adjustment is possible.  With the further requirement of radiative equilibrium, they depict the adjustment PDF parameter space for a power law opacity.  They find that with this further requirement, the adjustment PDF $\it{is~positive}$.  Our numerical results here confirm Baes et al.'s (2005) discussion along these lines, i.e., that the further requirement of radiative equilibrium allows BW01's method to produce correct results for any generic geometry.  The difference between Lucy's (1999) and BW01's method then boils down to a matter of practical importance - in large three-dimensional grids, there are many more path lengths (than in 1-d grids), and as such, the efficiency of Lucy's iterative method becomes considerably higher than FDA type algorithms.  Path length estimators can be constructed for absorption and emission by gas (Lucy 2003); iterative methods can be employed when opacities depend on temperature as in photoionization models (Ercolano et al. 2003), and when temperature fluctuations from small grains above equilibrium values are relevant for the emergent spectrum, i.e. for PAH emission, which we investigate in a future paper.    

In summary, we find that the Lucy method is the most efficient method for use in 3-D grids; we summarize the most optimal way of using the Lucy method given our findings regarding $N_{\rm temp}$ and $N_{\rm SED}$.  We find that $N_{\rm temp,per~iter} \sim N_{\rm grid}$ when using the fiducial $V4$ Lucy method yields a converged temperature profile to within 2 \%.  Typically four iterations are sufficient, i.e, $N_{\rm temp,total} \sim 4 N_{\rm grid}$.  (Note that this is a rule of thumb - in practice, we specify the tolerance on the standard deviation of the temperatures and not the number of iterations)  The number of photons used per iteration will also influence the accuracy of the longwave SED if the temperature is not converged.  Therefore, it is essential that this criterion be satisfied.  To get a noise-free SED, one can use a larger number of photons in the last iteration (once the temperatures are converged).  A clean long wave inclination averaged SED can be obtained with $N_{\rm SED} \sim 5 N_{\rm temp,per~iter}$.  As shown above, if the user is satisfied with a low level of noise at submillimeter wavelengths over 200 viewing angles, then $N_{\rm SED} \sim 24 N_{\rm temp,per~iter}$ is sufficient.  For the submillimeter and millimeter wavelengths, it is in fact acceptable to average over viewing angles since this part of the spectrum is not subject to the effects of optical depth.  As such, variations along viewing angle in this part of the spectrum reflect photon noise and not any intrinsic variations. 
The following equation summarizes our findings - 
\begin{equation}
N_{\rm total} \sim 4 N_{\rm temp,per~iter} + N_{\rm SED} \;.
\end{equation}

\subsection{SEDs of Multi-Dimensional Inhomogeneous Envelopes}

\begin{figure}[!ht] \begin{center}
\centerline{\psfig{file=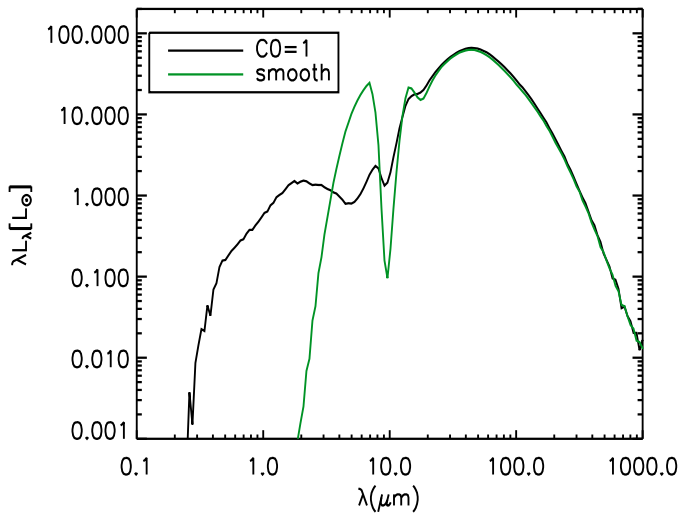,height=3.in,width=3.5in}}
\end{center}
\caption{Emergent SED of a clumpy envelope, $C_{0}=1$, (shown here averaged over 200 viewing angles) compared to SED of a homogeneous envelope where the density is smoothly varying, using the fiducial Lucy iterative method.}
\end{figure}

\begin{figure}[!hb] \begin{center}
\centerline{\psfig{file=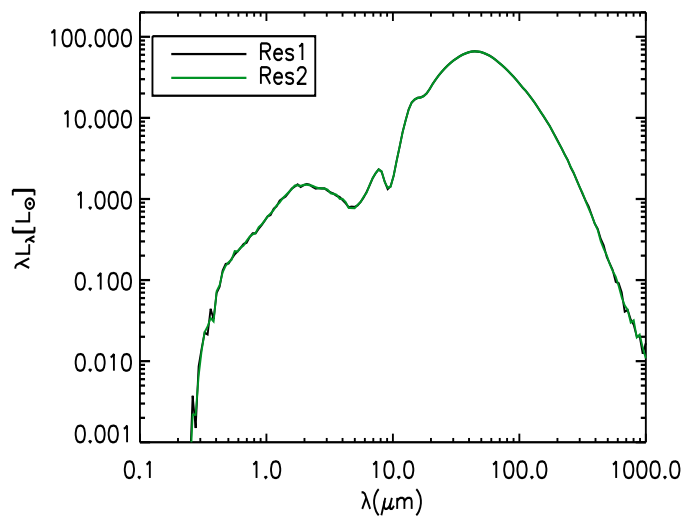,height=3.in,width=3.5in}}
\end{center}
\caption{Resolution study for the clumpy $C_{0}=1$ envelope.  The Res1 resolution is the standard resolution.  Increasing the resolution beyond this by a factor of 8 (2 times higher resolution in each dimension) does not affect the results.  SEDs are shown averaged over 200 viewing angles.}
\end{figure}

We now proceed to examine in detail the emergent SEDs of inhomogeneous or clumpy envelopes in three dimensions.  The envelopes have the same L,M, and R as before (luminosity, mass, and size) in keeping with CM05's parameterization of SEDs, so as to easily compare the emergent SED of a clumpy envelope with that of a homogenous envelope.  We parametrize the level of clumpiness in the envelope following Chakrabarti \& McKee (2008, in preparation; henceforth CM08), from which we quote a basic formula for the covering factor and parameterization of the clumpiness here.  (A somewhat different way of prescribing the level of clumpiness in inhomogeneous envelopes has been used by Indebetouw et al. 2006, and we compare with their prescription below as well) The covering factor, $\mathcal{C}$, is the number of clumps encountered along a line of sight:
\beq
\mathcal{C}=\int d\mathcal{C} =\int_{r_{min}}^{R_{c}} 4 \pi r^{2} n_{\rm cl}(r) \frac{\pi a_{\rm cl}^{2}(r)}{4\pi r^{2}} dr  \;,
\eeq
where $n_{\rm cl}(r)$ is the number density of clumps, and $a_{\rm cl}(r)$ is the size of the clumps, both of which may vary with radius.  $R_{c}$ and $r_{min}$ are the maximum and minium radius of locations of the clumps.  CM08 show that a useful form of the covering factor for a randomly spherically symmetric but clumpy envelope (i.e., one in which clumpiness is a function of radius and not angle) is given by: 
\beq
\mathcal{C}=\frac{3}{4}f_{\rm cl}q^{-1}\rm ln(\rct/\tilde{r}_{\rm min})~~~~for~a_{\rm cl}(r)=qr \;,
\eeq                               
where $f_{\rm cl}$ is the volume filling fraction, $\rct$ the ratio of the outer radius to the characteristic radius or Rosseland photosphere (e.g., CM05), and $\tilde{r}_{\rm min}$ is the ratio of the inner radius to the characteristic radius.  We can then identify as $\mathcal{C}_{0}$ the term $\frac{3}{4}f_{\rm cl}q^{-1}$, to which the covering factor is proportional.  Clumpy envelopes have low(er) volume filling fraction than homogeneous envelopes, i.e., for the same L,M, and R, they are characterized as having lower $\mathcal{C}_{0}$.  A clumpy envelope tends to a homogeneous envelope in the limit of large $\mathcal{C}_{0}$ (CM08). 

Figure 11 depicts the emergent SED averaged over 200 viewing angles of a clumpy envelope with $\mathcal{C}_{0}=1$ computed with the fiducial Lucy method, along with the emergent SED of a homogeneous envelope with the same L,M, and R as the clumpy envelope.  
One basic difference between a clumpy envelope and its equivalent homogeneous envelope (i.e., with the same L,M, and R but with the mass distributed in clumps) is that the effective optical depth of the clumpy envelope is lower.  This aspect of clumpy envelopes has been discussed by previous authors (Bowyer \& Field 1979; Varosi \& Dwek 1999; Slavin et al. 2000).  This can then lead to more high frequency flux escaping from the envelope, particularly at wavelengths where the optical depth of the homogeneous envelope was large.  However, the long wavelength flux is still optically thin, and as such we would expect the long wavelength flux for a clumpy envelope and its equivalent homogeneous envelope to be the same (i.e., as long as the long wavelength flux is optically thin).  Figure 11 indeed shows that this is the case.  The smooth and clumpy envelopes have nearly identical SEDs for $\lambda \ga 30~\micron$, but the clumpy envelope allows for high frequency (near-IR) flux to escape more easily.  Parts of the near-IR SED of the clumpy envelope are higher.  For instance, the $10~\micron$ region of the SED is higher when the envelope is clumpy.  The dust opacity curve has a silicate feature which leads to a maximum in the opacity at this wavelength.  However, when the envelope is clumpy, the effective optical depth through the envelope is lower, and it is easier for high frequency photons to escape.   An explanation of the essential differences of SEDs of clumpy and homogeneous envelopes is given in a forthcoming work (CM08).  Indebetouw et al. (2006) have also explored the differences in the emergent SEDs of clumpy and homogeneous envelopes and find similar results.  They emphasize that the emergent SED is inclination-dependent, particularly for the shorter wavelengths, which are subject to the effects of varying optical depths along different lines of sight.

\begin{figure}[!ht] \begin{center}
\centerline{\psfig{file=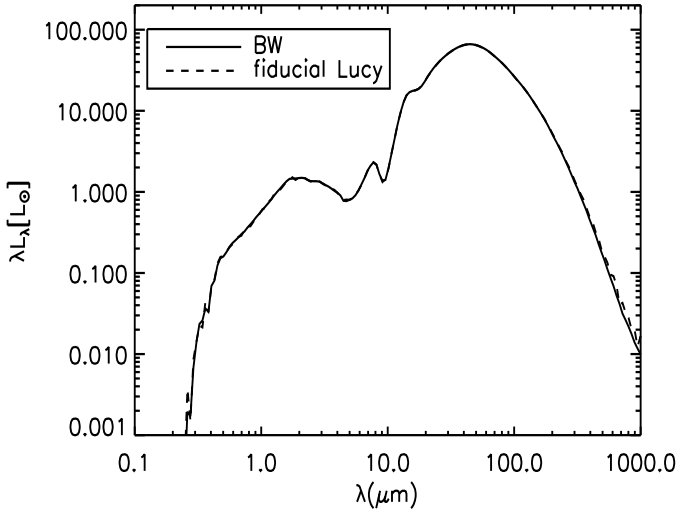,height=3.in,width=3.5in}}
\end{center}
\caption{Comparison of the SEDs averaged over 200 viewing angles calculated by the BW01 and fiducial Lucy methods for the $C_{0}=1$ envelope.}
\end{figure}

\begin{figure}[!ht] \begin{center}
\centerline{\psfig{file=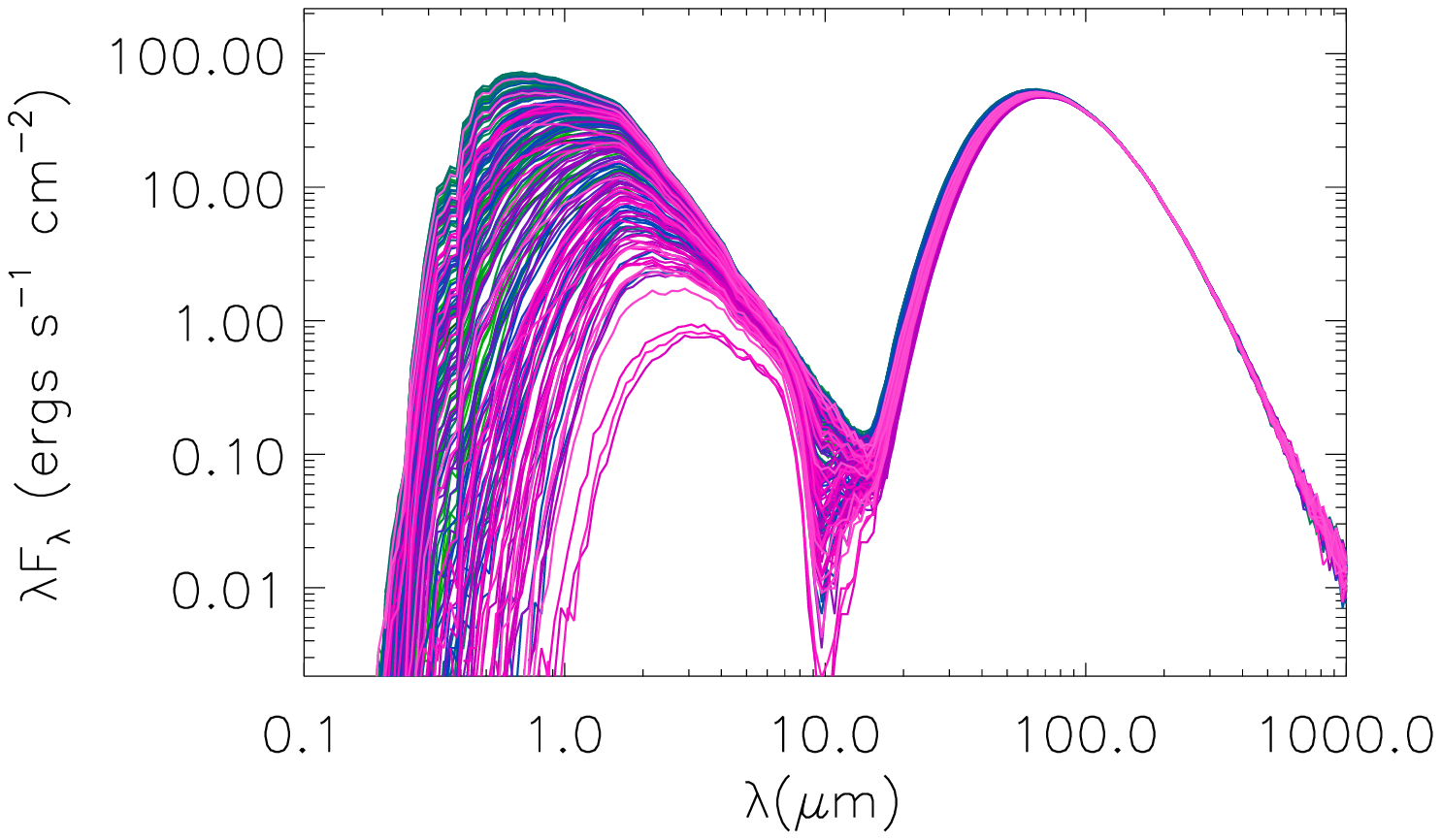,height=2.in,width=2.in}
\psfig{file=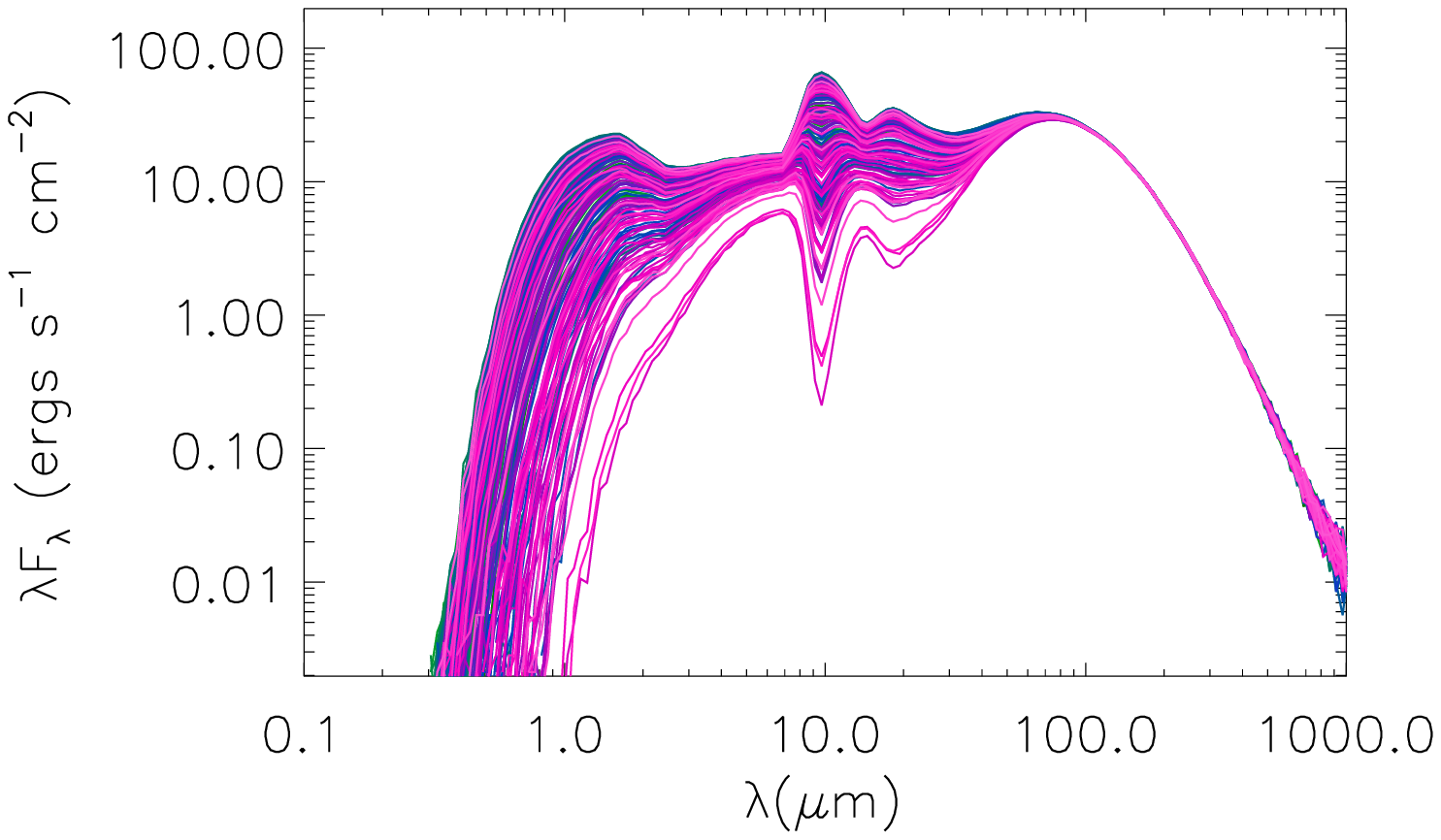,height=2.in,width=2.in}}
\end{center}
\caption{(a) Emergent SED over 200 viewing angles with no interclump diffuse density (b) Same as (a), but with a mass fraction in the diffuse interclump region that is 5 \% of the total mass.}
\end{figure}

Figure 12 demonstrates that our standard resolution case, denoted $Res1$, is sufficient to capture the inhomogeneities of this clumpy envelope.  Increasing the resolution by a factor of 8 (denoted $Res2$, 2 times in each dimension) does not alter the results.  Finally, Figure 18 compares the resultant SED of this clumpy envelope using the BW01 and fiducial Lucy iterative methods.  As is clear, both methods yield the same SED.  The fiducial iterative method required 2196.20 of CPU time, while BW01's method required 19243.76 of CPU time, i.e., iterating with the Lucy method is 9 times faster.     
However, these spherically averaged SEDs require less computing time.  
To produce comparable SEDs at each of 200 viewing angles requires 200 times more photons, as we show next.

Finally, we show in Figures 14 the emergent SED over 200 viewing angles including a diffuse interclump density, which is set to zero in Figure 14a and set to 5 \% of the total mass in Figure 14b.  Addition of this diffuse interclump phase does not affect the far-IR region of the SED ($\lambda \ga 30~\micron$) but significantly increases the amount of mid-infrared emission along some lines of sight in clumpy envelopes.  These models use a fractal density distribution (Whitney et al. 2005,
Indebetouw et al. 2006) with a fractal dimension of D=2.3 chosen to match
that of the ISM (Elmegreen 1997).  
500 million photons were run through the final iteration to produce clean SEDs over all viewing angles.

\section{RADISHE: Radiative Transfer Through Hydrodynamical Simulations of Dusty Galaxies}

\begin{figure}[!hb] \begin{center}
\centerline{\psfig{file=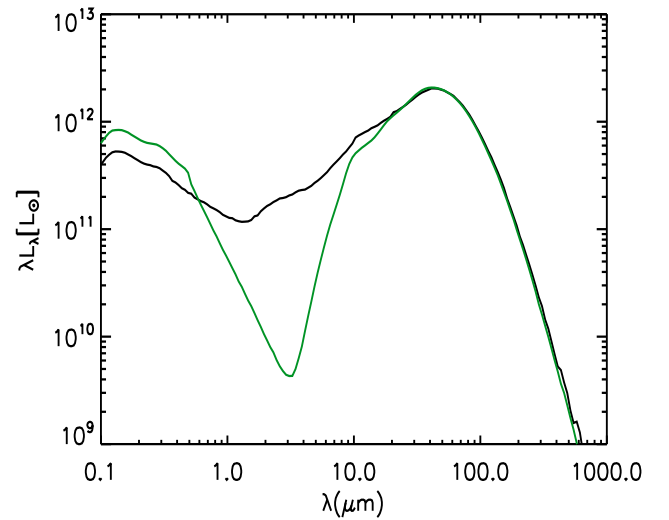,height=3.in,width=3.5in}}
\end{center}
\caption{The emergent SED shown averaged over 200 viewing angles during a quasar phase (where $L_{\rm BH} \sim L_{\rm bolometric}$) from the A5e simulation, with (black line) and without (green line) the inclusion of the hot dust component in the AGN input spectrum.}
\end{figure}

Hydrodynamical codes that model the evolution of galaxies, like GADGET-2 (Springel 2005), calculate, among other variables, the time evolution of the gas densities, stellar masses, star formation rate, and the accretion rate onto the black hole.  All of these quantities are changing as a function of time, as gas is converted into stars, accreted by the black hole, or expelled by energetic feedback.  Thus, the SEDs calculated along a galaxy simulation time sequence, i.e., the calculated photometric history of these galaxies, directly reflects the evolutionary history of the simulated galaxy.  We describe below the basic setup for prescribing the dust distribution given the gas densities, and how distributed sources of radiation, like star particles, are treated in the code.  We have considered here a subset of the simulations described in Chakrabarti et al. (2007b) (we use the same naming convention as in that paper), which we found are representative of moderate to high redshift infrared bright galaxies.  

We focus our discussion on $z\sim 2$ galaxies since this where the quasar and SMG luminosity function peaks (Richards et al. 2006; Chapman et al. 2005), and $\it{Herschel}$ is expected to secure rest-frame observations close to the peak of the SEDs for large samples of infrared bright galaxies at $z \sim 2$.  We model SMGs as products of major mergers which undergo an intense period of feedback from AGN.  Local ULIRGs have been shown to be the products of major mergers (Dasyra et al. 2006).  However, the suggested formation mechanisms and evolutionary scenarios for SMGs remain varied in nature, ranging from primeval, heavily accreting galaxies undergoing a starburst (Rowan-Robinson 2000; Efstathiou \& Rowan-Robinson 2003) to products of gas-rich major mergers undergoing intense feedback (Chakrabarti et al. 2007b), with recent observations favoring the latter scenario (Nesvadba et al. 2007; Bouche et al. 2007).  In particular, Bouche et al. (2007) suggest
that dissipative major mergers may have produced the SMG population on the basis of their
finding that the SMG population has lower angular momenta and higher matter densities
compared to the UV/optically selected population.  We compare our calculated SEDs to data of SMGs studied by Pope et al. (2006), that span the range of evolutionary stages described in Chakrabarti et al. (2007b), i.e., from the Class I phase (where the stars dominate the bolometric luminosity, and the SMG has a high value of $L_{\rm IR}/L_{\rm x} \sim 100$) to the x-ray bright Class II phase where the black hole dominates the energetics of the galaxy ($L_{\rm IR}/L_{\rm x} \sim 25$).

The galaxy simulations we consider here are described in detail in Chakrabarti et al. (2007a;2007b) and Cox et al. (2007); we consider here specifically the simulations A3,A4,A5,A4e, and A5e, where ``e'' denotes a non co-planar merger and A3,A4,A5 have virial velocities of 160, 226, and 320 $\rm km/s$ respectively.  
GADGET-2 specific methodology is described in Springel et al. (2005) and Springel (2005).  We give a brief description here.  We employ a new version of the parallel TreeSPH code GADGET-2
(Springel 2005), which uses an entropy-conserving formulation of
smoothed particle hydrodynamics (Springel \& Hernquist 2002), and
includes a sub-resolution, multiphase model of the dense interstellar
medium (ISM) to describe star formation (Springel \& Hernquist 2003,
henceforth SH03).  The multiphase gas is pressurized by feedback from
supernovae, allowing for stable evolution of even pure gas disks. 
Black holes are represented by ``sink''
particles that accrete gas, with an accretion rate estimated using a
Bondi-Hoyle-Lyttleton parameterization, with an upper limit equal to
the Eddington rate (Springel et al. 2005b). The bolometric luminosity
of the black hole is then $\L_{\rm bol}=\epsilon_{\rm r}\dot{M}c^{2}$,
where $\epsilon_{\rm r}=0.1$ is the radiative efficiency.  We further
allow a small fraction ($\sim 5\%$) of $\L_{\rm bol}$ to couple dynamically
to the gas as thermal energy. This fraction is a free parameter,
determined in Di Matteo et al. (2005) by matching the $M_{\rm
BH}$-$\sigma$ relation.  We do not attempt to resolve the gas
distribution immediately around the black hole, but instead assume
that the time-averaged accretion can be estimated from the gas on the
scale of our spatial resolution, which is typically $\sim \rm 50~pc$.

RADISHE is based on the code described in Whitney et al. (2003); 
the primary algorithmic differences are the temperature algorithm used in RADISHE 
which is the iterative Lucy method (1999) that has been described in \S 2-3, and the use of 
distributed sources of radiation; SPH specific interface aspects, such as the input densities and 
dust-to-gas ratios for various phases of the ISM are described below.

\subsection{Basic Setup \& Convergence Studies} 

We adopt a ISM prescription similar to that developed by Chakrabarti et al. (2007a;2007b)
in their study of the infrared properties of local ULIRGs.
Owing to the limited resolution of the SPH simulations, the
interstellar medium (ISM) is tracked in a volume-averaged manner.
In particular, the simulations discussed here adopt the multiphase
model of Springel \& Hernquist (2003; henceforth SH03), in which individual SPH particles represent a region of
the ISM that contains cold clouds embedded in a diffuse hot medium.
Because the SPH calculation uses only the volume-averaged pressure,
temperature, and density to evolve the hydrodynamics, this model does
not provide specific information regarding the distribution (i.e., masses, sizes) of the clumps.  
As such, we adopt a sub-resolution model guided by observations of GMCs.  Observations of GMCs 
indicate that power-law scaling relations (Larson 1981) describe the cloud
velocity dispersion and size.  GMCs have a distribution of masses ($dN/dM \propto M^{-\alpha}$)
 (Williams \& McKee 1997; Rosolowsky 2005; Blitz et al. 2007) where $\alpha \sim 1.8$ 
such that most clouds are on the low mass tail, with most of the molecular mass being contained
in the most massive clouds.  The mass-radius relation ($M \propto R^{\gamma}$) of GMCs is found to 
have an average value of $\gamma \sim 2$ (Rosolowsky 2005; Rosolowsky 2007; Blitz et al. 2007), 
which we adopt in this work.  Observations of molecular clouds indicate that their pressure is
dominated by turbulent motions (Blitz et al. 2006; Solomon et al. 1997), rather than thermal energy, and thus
neglecting this feature yields very dense clouds, with small volume
filling factors.  Including turbulent support is also motivated by recent models star formation (Krumholz \& McKee 2005) that are able recover the Kennicutt-Schmidt relation. In the work presented here, we take the turbulent
pressure to be $\sim 100$ times that of the thermal pressure; the turbulent
velocity as estimated from line widths is of order 10 times the thermal
sound speed in star forming regions, which would lead to a factor of $\sim 100$ 
for the ratio of the turbulent to thermal pressure (Plume et al. 1996).  With this
assumption, the density is much lower than without the turbulent support
and its volume filling factor is greatly increased.  (Our approach is similar to that employed
by Narayanan et al. [2006] in their study of the evolution of
the molecular gas in mergers.)  We take the diffuse, hot gas ($\ga 10^{4} \rm~K$) to be uniformly
 distributed. 

Dust formation and destruction is not tracked in the hydrodynamical simulations.  
We assume that the dust spatial distribution follows the gas, i.e., we assume
that the dust and gas are coupled.  
We adopt a dust-to-gas ratio for the diffuse phase that is
one-tenth of that in the dense phase, as we empirically find that this value fits the observed inclination averaged SEDs of SMGs (the SED does vary significantly with inclination and higher or lower dust-to-gas ratios may be favored for face-on or edge-on inclinations.  This particular value is suited to simulations of gas-rich systems as we have considered here; we perform a more extensive parameter study in a forthcoming paper).  We adopt the dust model of Weingartner \& Draine (2001) $R_{\rm V}=5.5$; the dust-to-gas ratio adopted for the dense phase is the same as in WD01,
i.e., 1/105.1 for solar abundances, with a factor of two increase to account for ice mantles (see also CM05, Chakrabarti \& McKee 2007, and Chakrabarti et al. 2007a;b for a discussion of this).  From observations of their far-IR SEDs, Dunne \& Eales (2001) and Klaas et al. (2001) found that the dust-to-gas ratio for a large sample of ULIRGs was comparable to Milky Way values, when they fit two-temperature blackbodies to the far-IR SEDs.  Use of a self-consistent analytic radiative transfer solution with a continuous range of temperatures also fits the far-IR SEDs of a large sample of local ULIRGs when the WD01 dust opacity, with ice mantles, is adopted (Chakrabarti \& McKee 2007).

We have performed Starburst 99 calculations (Leitherer et al. 1999; Vazquez \& Leitherer 2005) to calculate the stellar spectrum and bolometric stellar luminosity, taking as input the age, mass, and metallicity of the stars from the simulations, for a Salpeter IMF.  Due to the large number of distributed sources of radiation in the simulations (sometimes of order $\sim 10^{5}$), after we obtain the stellar spectrum, we approximate the stellar spectrum as a blackbody at some effective temperature.  In other words, we sample the Planck function with the appropriate $T_{\star}$ for every star particle.  This affects our results only at very short wavelengths ($\lambda < 0.1 ~\micron$), where the stellar spectrum deviates from a blackbody.  Photons are emitted simultaneously from both the nuclear source of radiation (the black hole) and these distributed sources of radiation (stars).  The off-nuclear sources of radiation are allocated photons in proportion to the fraction of bolometric luminosity they contribute.  As we discuss in more detail later in the context of grid resolution, we do not here resolve the sharp temperature increase that occurs near the surface of a star (e.g. Ivezic \& Elitzur 1997).  The resolution of the SPH simulations that we analyze is $\sim 50 \rm ~pc$ and therefore firstly a ``star particle'' really corresponds to a star cluster.   Secondly, since the simulations do not robustly track the time evolution of the gas densities on scales less than $\sim 50 \rm ~pc$, there is some inherent ambiguity as to how to specify the densities on a spatial grid for scales less than this (see Stamatellos \& Whitworth 2005 for an alternative approach where concentric shells are used to resolve the temperature structure around star particles).  We defer a full exploration of the details of the temperature structure close to star particles to a future paper.

We model the intrinsic AGN continuum spectrum following Hopkins, Richards
\& Hernquist (2006) (HRH), which is based on optical through hard X-ray observations
(Elvis et al. 1994, George et al. 1998, Perola et al. 2002, Telfer et
al. 2002, Ueda et al. 2003, Vignali et al. 2003), with a reflection
component generated by the PEXRAV model (Magdziarz \& Zdziarski 1995).
The HRH spectrum is similar to that developed by Marconi et al. (2004), 
but it is more representative of the shapes of typical observed quasars;
it includes a template for the observed ``hot dust'' component
 longward of $1~\micron$.  This ``hot dust'' originates from the dust torus
 surrounding the AGN on scales of $\sim \rm pc$ (Antonucci 1993; Honig et al. 2006; Kishimoto et al. 2007).  Since the simulations we analyze do not resolve 
the structure close to the AGN, we include this hot dust component in the AGN input spectrum,
which we scale to the luminosity of the black hole.  Figure 15 shows the emergent
spectrum, averaged over 200 viewing angles, of a simulation where the quasar dominates the bolometric luminosity 
with and without the inclusion of this hot dust component.  As is clear, including
this component leads to the near-IR power law (or approximately flat in $\lambda L_{\lambda}$) SED characteristic of energetically active AGN (Lacy et al. 2004; Richards et al. 2006).  High resolution SPH simulations may eventually be able to follow the time evolution of the gas densities on these small scales and self-consistently produce this feature; our current treament is analogous to a sub-grid (or sub-resolution) specification of this component.

\begin{figure}[!ht] \begin{center}
\centerline{\psfig{file=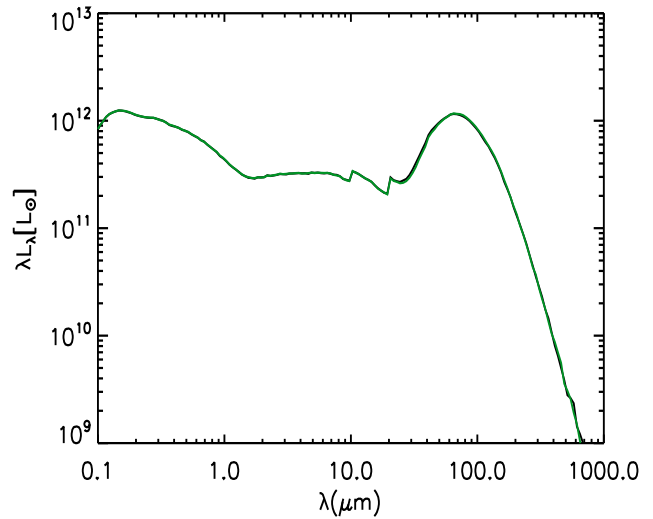,height=3.in,width=3.5in}}
\end{center}
\caption{Emergent SED shown averaged over viewing angles as the grid resolution is varied by a factor of 8 (the black is the higher resolution and the green is the lower resolution), for a timesnap close to the quasar phase in the A5e simulation.}
\end{figure}

Figure 16 shows the emergent SED, averaged over 200 viewing angles, of a quasar-like timesnap (when the black hole dominates the bolometric luminosity) from the A5e simulation for two different grid resolutions.  The black line and green lines depict SEDs from identical simulations, with one run at eight times higher resolution (black line).  Our standard resolution is the one shown in green; as is clear, increasing the grid resolution does not alter the results.  Our fiducial grid resolution corresponds to $a_{\rm cl} \sim 30 \Delta r~$ (where $a_{\rm cl}$ is the median size of the cold clump size distribution) close to the Rosseland photosphere, which is on scales of a couple of hundred parsecs typically, and $a_{\rm cl} \sim 2-3 \Delta r~$ in the outer regions of the galaxy, i.e., out at $\sim 10~\rm kpc$.  (To avoid over-resolving the diffuse phase relative to the smoothing length of the simulations, we do not include a diffuse phase component for radii less than the minimum smoothing length) This level of grid resolution is more than sufficient at capturing density inhomogeneities that the hydrodynamical simulation is able to follow. As discussed before, we have not attempted here to capture the sharp temperature increase close to star particles, which occurs on scales much smaller, i.e., $ \sim$ few  $\rm pc$, than what is probed by these hydrodynamical simulations.  

\begin{figure}[!ht] \begin{center}
\centerline{\psfig{file=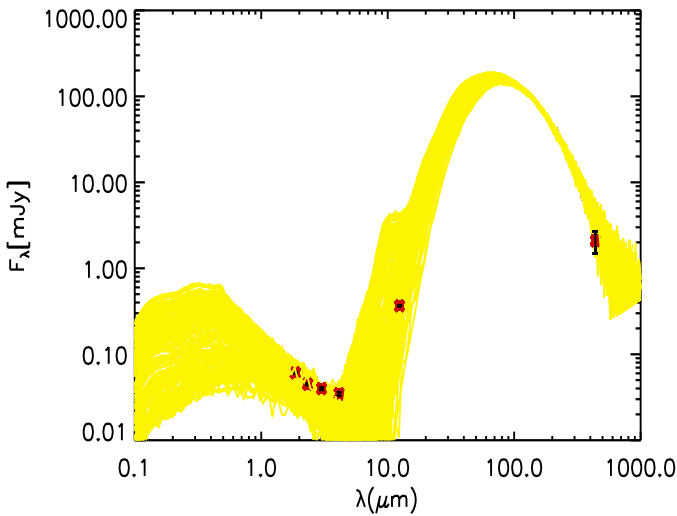,height=3.in,width=3.5in}}
\end{center}
\caption{The rest-frame emergent SED during the Class I phase from the A5e simulation, shown here over 200 viewing angles, compared to data of a SMG at $z=0.935$ (denoted GN31 in Pope et al. (2006)).}
\end{figure}

\begin{figure}[!ht] \begin{center}
\centerline{\psfig{file=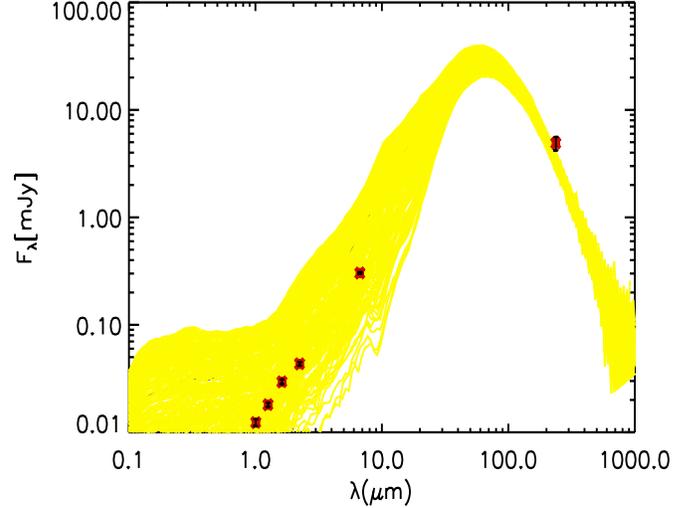,height=3.in,width=3.5in}}
\end{center}
\caption{The rest-frame emergent SED during the Class II phase from the A5e simulation, shown here over 200 viewing angles, compared to data of a SMG at $z=2.578$ (denoted GN04 in Pope et al. (2006)).}
\end{figure}

\subsubsection{The Class I $\rightarrow$ Class II Transition in SEDs}

Figure 17 depicts the emergent SED of an evolutionary phase from the A5e simulation where the light from the stars dominates the bolometric luminosity, with a star formation rate of $300 M_{\odot}/\rm yr$ and the black hole providing 30 \% of the total luminosity.  The calculated SED is compared to SCUBA and $\it{Spitzer}$ data of a SMG at $z \sim 0.9$ from Pope et al. (2006).  We termed this phase of evolution the Class I phase in Chakrabarti et al. 2007b - this is phase of the evolution close to the final merger where the stars dominate the bolometric luminosity and $L_{\rm IR}/L_{\rm x} \sim 100$.  Figure 18 depicts a comparison of the emergent SED during the Class II phase of the A5e simulation, where the black hole contributes $\sim 80$\% to the total luminosity, to $\it{Spitzer}$ and SCUBA data of a $z \sim 2$ X-ray selected SMG discused in Pope et al. (2006).  Pope et al. (2006) note in their discussion of this source that the morphology of this source is suggestive of a merger origin.  Of the candidate AGN they discuss, they find that the AGN in this source very likely dominates the observed-frame mid-IR emission.  The power-law mid-IR SED seen here is characteristic of energetically active AGN (Richards et al. 2006).  We termed this phase of evolution the Class II phase in Chakrabarti et al. 2007b; it is characterized by high intrinsic hard X-ray luminosities ($L_{\rm x} \ga 10^{11}~L_{\odot}$) and $L_{\rm IR}/L_{\rm x} \sim 25$.  Sources in this phase of evolution are transitioning between the infrared bright SMG phase (which spans the Class I-II stages) to infrared dim merger remnants following a brief blowout phase when feedback from the AGN is effective at dispersing the columns of dust and gas within the galaxy.  In a forthcoming paper, we present results from an automated search via a web interface to fit observed SEDs of infrared bright galaxies, in a similar vein as the work on analysis of YSO SEDs performed by Robitaille et al. (2006).

\section{Correlating $\it{Herschel}$ Color-Color Plots With Galactic Energy Sources}

\begin{figure}[!ht] \begin{center}
\centerline{\psfig{file=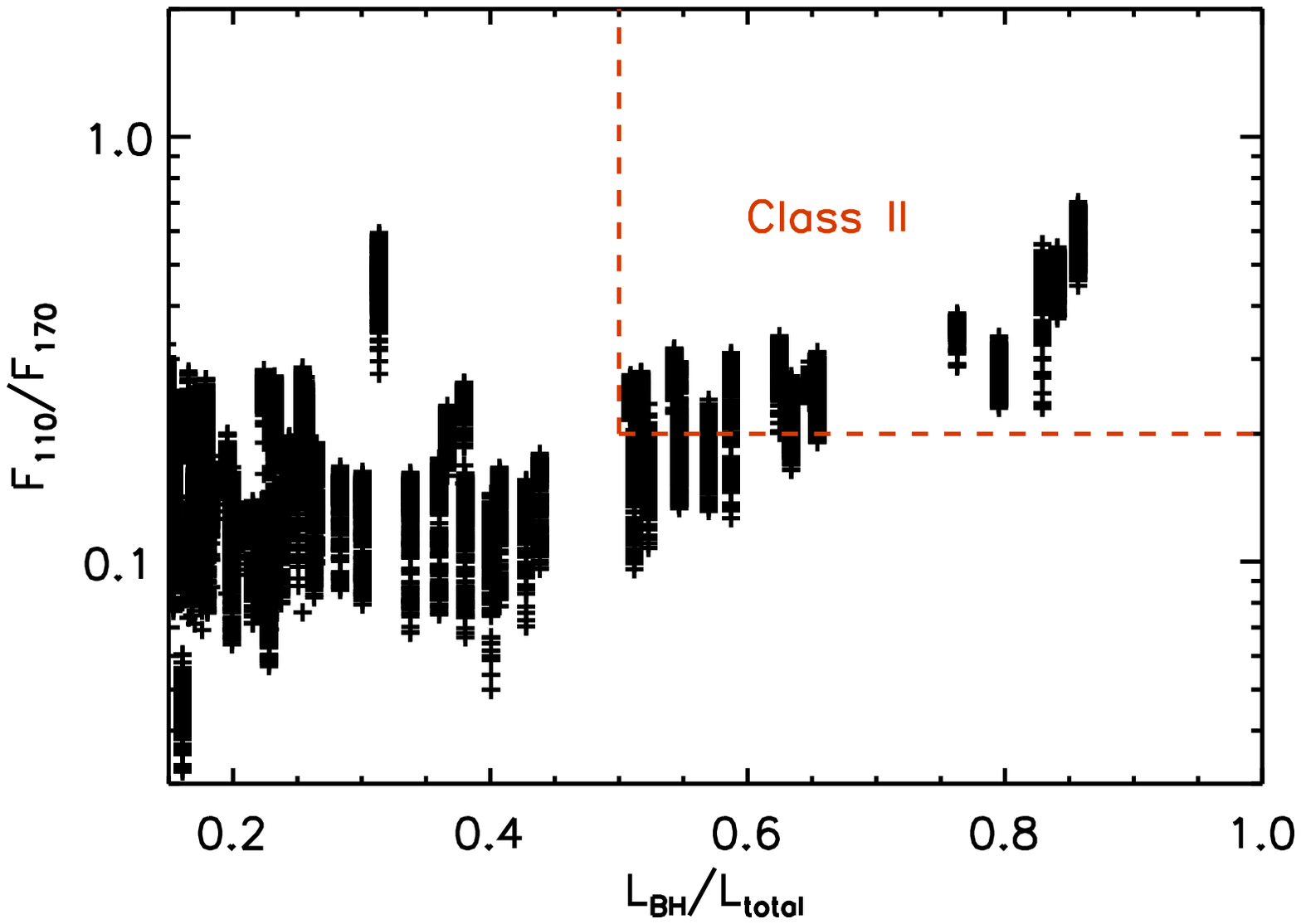,height=2.in,width=2.in}
\psfig{file=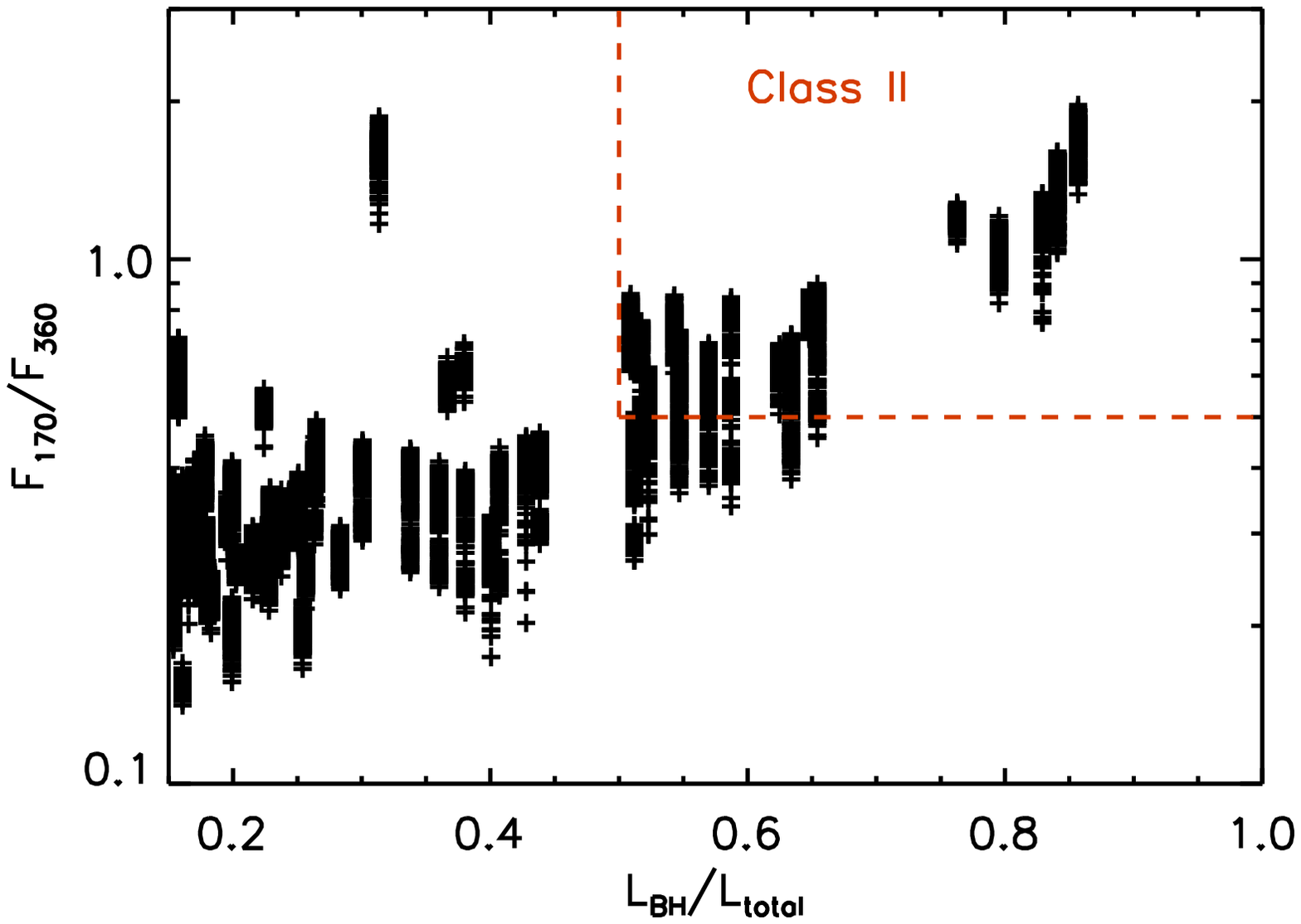,height=2.in,width=2.in}}
\end{center}
\caption{Observed frame color ratios for $z=2$ galaxies as a function of $L_{\rm BH}/L_{\rm total}$ for several simulations.  (a) This is rest-frame $F_{\rm 36.6~\micron}/F_{56.6~\micron}$ so these bands probe the cold-warm trend; see text for discussion.  High $F_{\rm 110~\micron}/F_{\rm 170~\micron}$ is correlated with energetically active AGN. (b)High $F_{\rm 170~\micron}/F_{\rm 360~\micron}$ is also correlated with energetically active AGN.  The Class II phase is demarcated in a conservative manner - to encompass the color region that is always (even given viewing angle dependences) correlated with $L_{\rm BH} \ga 0.5 L_{\rm total}$.}
\end{figure}

 \begin{figure}[!ht] \begin{center}
\centerline{\psfig{file=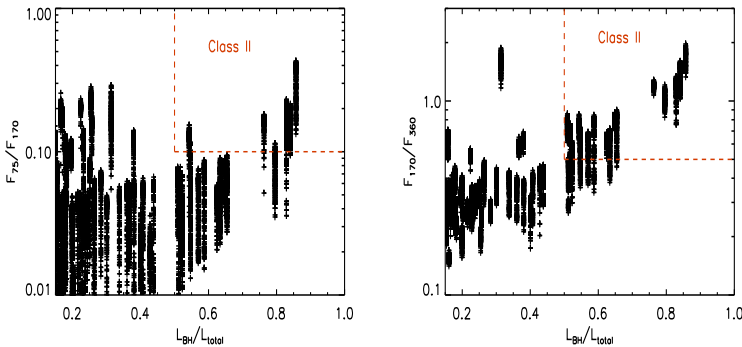,height=2.in,width=2.in}
\psfig{file=F170_360_z2_LBHLtot_A5eA4eA3A5A4_inc_nocol.ps,height=2.in,width=2.in}}
\end{center}
\caption{Observed frame color ratios for $z=2$ galaxies as a function of $L_{\rm BH}/L_{\rm total}$ for several simulations.  (a) This is rest-frame $F_{\rm 25~\micron}/F_{56.6~\micron}$; high $F_{\rm 110~\micron}/F_{\rm170~\micron}$ is correlated with energetically active AGN. (b)High $F_{\rm 170~\micron}/F_{\rm360~\micron}$ is also correlated with energetically active AGN.}
\end{figure}

A significant result from the IRAS survey was the robust identification of energetically active AGN in local ULIRGs on the basis of IRAS colors, specifically the coincidence of high $25~\micron$ to $60~\micron$ flux ratios (or ``warm'' sources) (de Grijp et al. 1985; Sanders \& Mirabel 1996) and energetically active AGN.  It was also found that higher ratios of $F_{\rm 60~\micron}/F_{\rm 100~\micron}$ were also correlated with energetically active AGN (Soifer et al. 1987).  $\it{Herschel}$ is expected to probe close to these rest-frame colors for $z \sim 2$ galaxies, which has not been previously possible due to lack of sensitivity.  We have previously shown in an earlier paper that ``warm'' $25~\micron/60~\micron$ colors can be reproduced and explained on the basis of AGN feedback (Chakrabarti et al. 2007a).  Specifically, we showed that $F_{25~\micron}/F_{60~\micron} \ga 0.15$ corresponds to the evolutionary phase in the simulations when $L_{\rm BH} \sim L_{\rm total}$.  As noted in Di Matteo et al. (2005) and Springel et al. (2005), the evolutionary phase where $L_{\rm BH} \sim L_{\rm total}$ is characterized by a deposition of thermal energy into the surrounding gas to recover the $M_{\rm BH}-\sigma$ relation.  A key aspect of this phase of evolution and its impact on emergent SEDs that had not been previously emphasized in the literature is that it is in this phase of evolution that the energetic feedback from the AGN is effective at clearing out the obscuring columns of dust and gas, thereby lowering the column of dust through the envelope.  Both the increase in luminosity (or $L/M$ in the formalism of CM05) and the lower optical depth through the envelope (or $\Sigma$ in the formalism of CM05) serve to increase the source of high frequency photons and allow the high frequency photons to escape the dust envelope more easily.

We focus here on $z \sim 2$ galaxies, as Chapman et al. (2003; 2005) have found this to be the median redshift of the SMG population.  We show in Figures 19 and 20 the observed-frame $\it{Herschel}$ band colors $F_{110~\micron}/F_{170~\micron}$, $F_{170~\micron}/F_{360~\micron}$, and $F_{75~\micron}/F_{170~\micron}$ for all 200 viewing angles, as a function of the ratio of the black hole's luminosity to the total bolometric luminosity for several simulations which are representative of high-redshift massive star forming galaxies (these are the A3,A4,A4e,and A5e simulations from Chakrabarti et al. 2007b).  These colors correspond to rest-frame $F_{\rm 36.6~\micron}/F_{56.6~\micron}$, $F_{\rm 56.6~\micron}/F_{\rm 120~\micron}$ and $F_{25~\micron}/F_{56.6~\micron}$ respectively, so these $\it{Herschel}$ bands are particularly suited to probe the effect of AGN feedback for $z \sim 2$ galaxies.  Alexander et al. (2005) have analyzed deep $\it{Chandra}$ observations to find that many radio-selected SMGs host a buried AGN.  Nesvadba et al. (2007) have recently found particularly impressive examples of large-scale jets in radio-loud SMGs - highlighting the need to explore the effect of jets and outflows in the SMG population.  Therefore, it is now useful to develop diagnostics in $\it{Herschel}$ bands that correlate with feedback from energetically active AGN in SMGs.  

As is clear, high ratios of these colors are well correlated with $L_{\rm BH} \ga 0.5 L_{\rm total}$, marked by the dashed lines, and denoted ``Class II''.
The previous plots also demonstrate the important point that three-dimensional inhomogeneous envelopes, such as the ones we have modeled here, will have mid-infrared colors that vary with viewing angles, as these frequencies are sensitive to the distribution of optical depths and temperature which vary with lines of sight through the envelope.  (For simplicity, Chakrabarti et al. 2007a presented the photometric time evolution of galaxy merger simulations by averaging the emergent colors over viewing angles.)  It is clear from Figures 19 and 20 that the shorter wavelength color ($F_{75~\micron}/F_{170~\micron}$) has a larger dispersion that the longer wavelength color ($F_{110~\micron}/F_{170~\micron}$), which is to be expected.  Thus, it may be useful to perform wide-area surveys in both the $F_{110~\micron}/F_{170~\micron}$ and $F_{75~\micron}/F_{170~\micron}$ bands, rather than only the $F_{75~\micron}/F_{170~\micron}$ band, to robustly derive observational diagnostics for AGN activity that are not biased by viewing angle dependences.  Unification models of AGN (e.g. Antonucci 1993) have emphasized that the inhomogeneous geometry of AGN hosts leads to observed quantities varying significantly with orientation.  Three-dimensional models of dust envelopes provide a new avenue to investigate these effects, which was not incorporated in the first generation of axisymmetric models (Efstathiou \& Rowan-Robinson 1995; Silva et al. 1998).  Although there is a notable dispersion with viewing angle, the overall $\it{evolutionary}$ trend, i.e., the increase in $F_{110~\micron}/F_{170~\micron}$ with $L_{\rm BH}/L_{\rm total}$ is preserved.   (As Chakrabarti et al. (2007a) have noted, the early and late phases of the final coalescence phase of the merger can both be, for some simulations, characterized by ``cold'' colors; as such, the Class I and Class III phases cannot be clearly separated on the basis of rest-frame mid-IR to far-IR colors that $\it{Herschel}$ is expected to probe.)  

\begin{figure}[!ht] \begin{center}
\centerline{\psfig{file=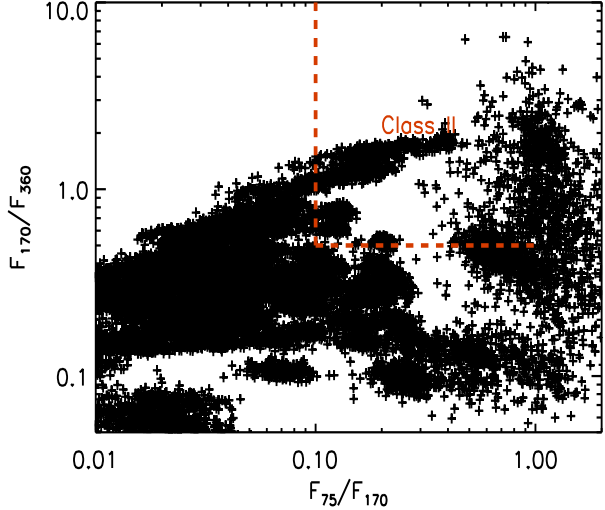,height=3.in,width=3.5in}}
\end{center}
\caption{Predictions for $\it{Herschel}$ color-color plots (in observed frame) for $z=2$ galaxies with the ``Class II'' evolutionary phase marked.  Having both of these colors leads to a greater constraint for identifying energetically active AGN than one color alone.  Galaxies tend to cluster in low regions of $F_{\rm 170~\micron}/F_{\rm 360}$ and low $F_{\rm 75~\micron}/F_{\rm 170~\micron}$ during most of their lifetime, and spend a small fraction of their time in high regions of the $F_{\rm 170~\micron}/F_{\rm 360}$-$F_{\rm 75~\micron}/F_{\rm 170~\micron}$ color-color plot.  During this phase (when $\it{both}$ colors are high) the AGN contributes significantly to the bolometric luminosity and is effective at expelling obscuring columns of dust and gas, lowering the effective optical depth through the envelope - thereby producing ``warm'' colors.}
\end{figure}

\begin{figure}[!ht] \begin{center}
\centerline{\psfig{file=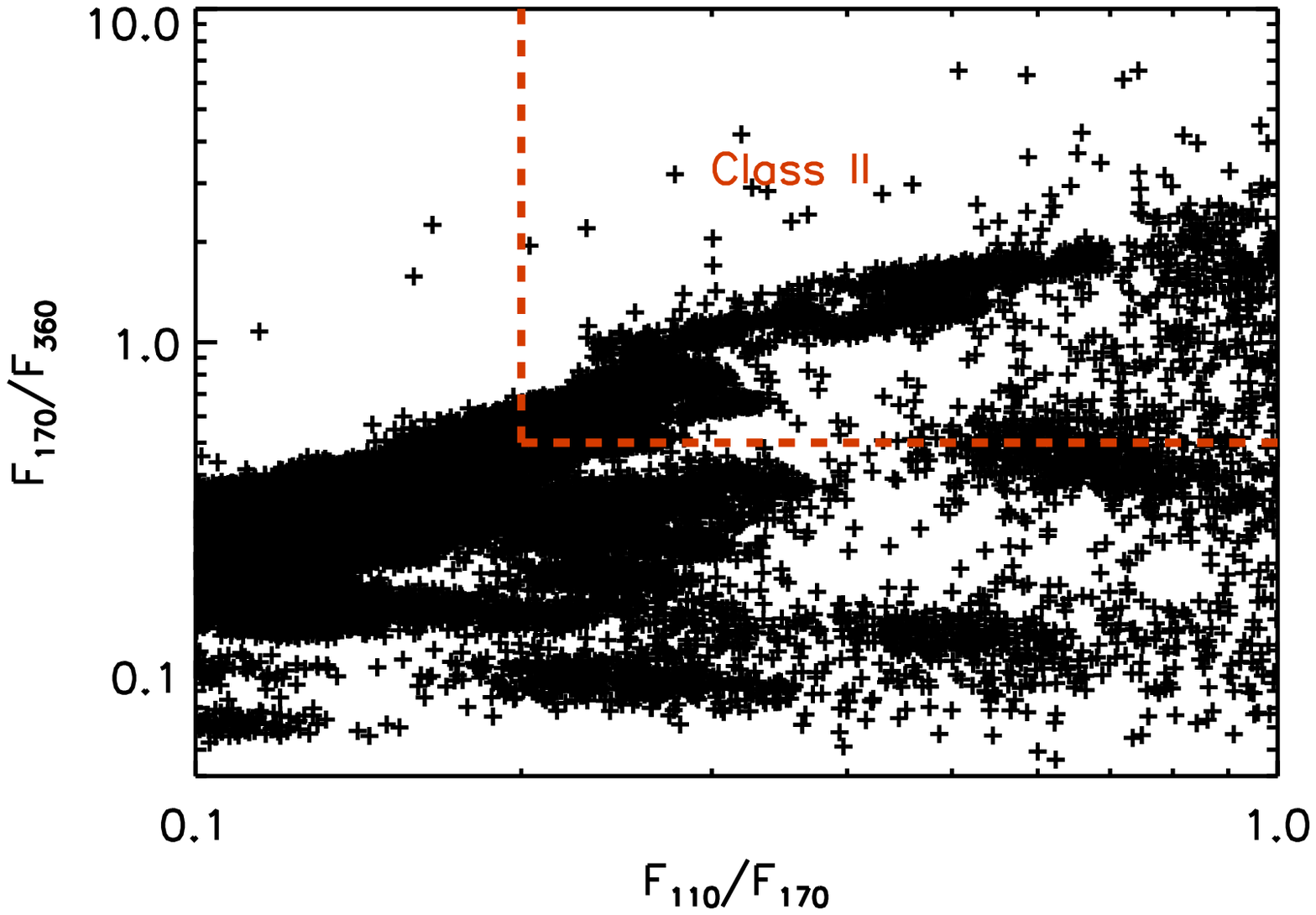,height=3.in,width=3.5in}}
\end{center}
\caption{Predictions for $\it{Herschel}$ color-color plots (in observed frame) for $z=2$ galaxies with the ``Class II'' phase marked.  Having both of these colors leads to a greater constraint for identifying energetically active AGN than one color alone.  Galaxies tend to cluster in low regions of $F_{\rm 170~\micron}/F_{\rm 360}$ and low $F_{\rm 110~\micron}/F_{\rm 170~\micron}$ during most of their lifetime, and spend a small fraction of their time in high regions of the $F_{\rm 170~\micron}/F_{\rm 360}$-$F_{\rm 110~\micron}/F_{\rm 170~\micron}$ color-color plot.  During this phase (when $\it{both}$ colors are high) the AGN contributes significantly to the bolometric luminosity and is effective at expelling obscuring columns of dust and gas, lowering the effective optical depth through the envelope - thereby producing ``warm'' colors.}
\end{figure}

\begin{figure}[!ht] \begin{center}
\centerline{\psfig{file=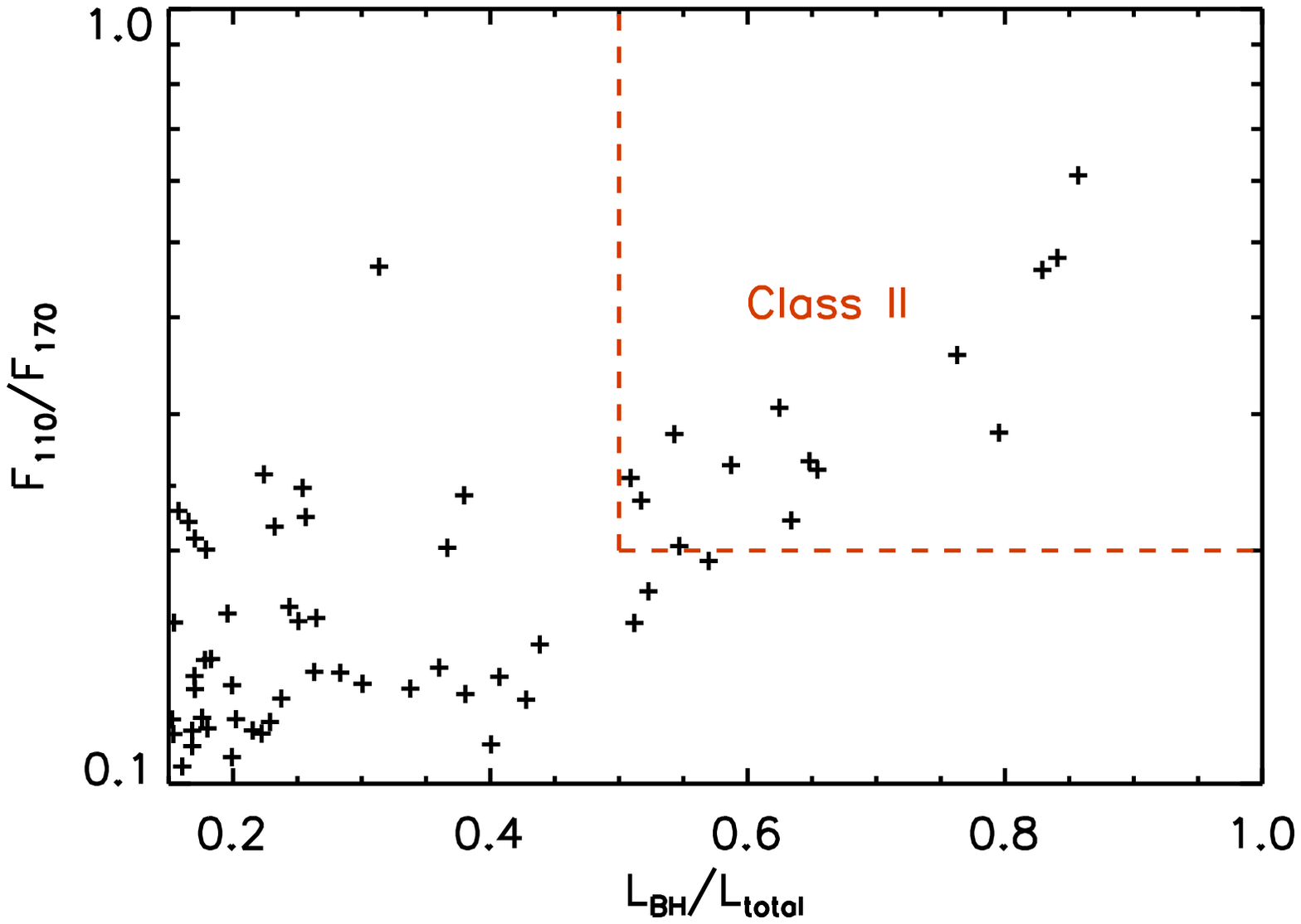,height=2.in,width=2.in}
\psfig{file=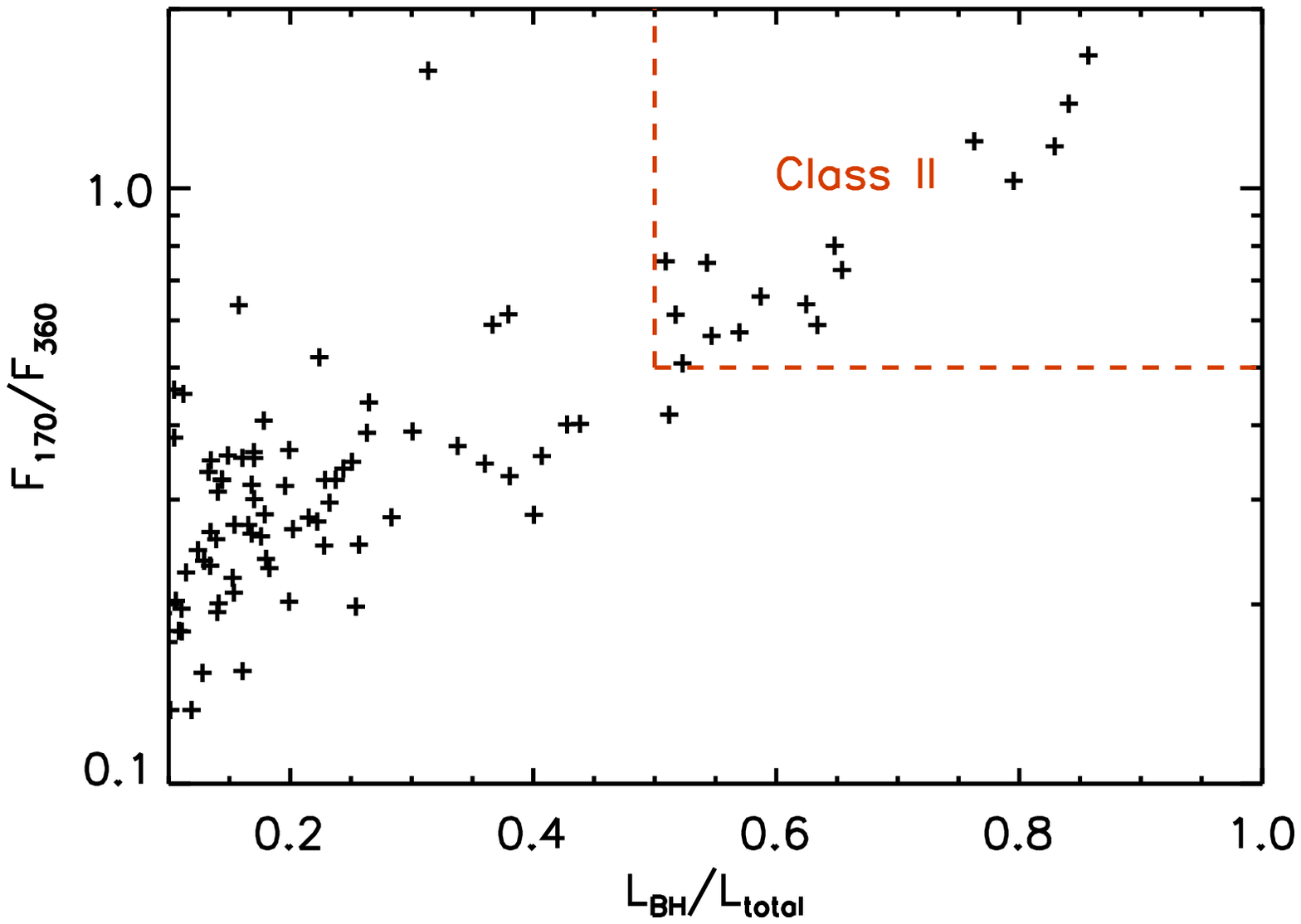,height=2.in,width=2.in}}
\end{center}
\caption{Observed frame color ratios for $z=2$ galaxies as a function of $L_{\rm BH}/L_{\rm total}$ for several simulations, shown here averaged over 200 viewing angles.  (a) High $F_{\rm 110~\micron}/F_{\rm170~\micron}$ is correlated with energetically active AGN as before, (b)High $F_{\rm 170~\micron}/F_{\rm360~\micron}$ is correlated with energetically active AGN. }
\end{figure}

\begin{figure}[!ht] \begin{center}
\centerline{\psfig{file=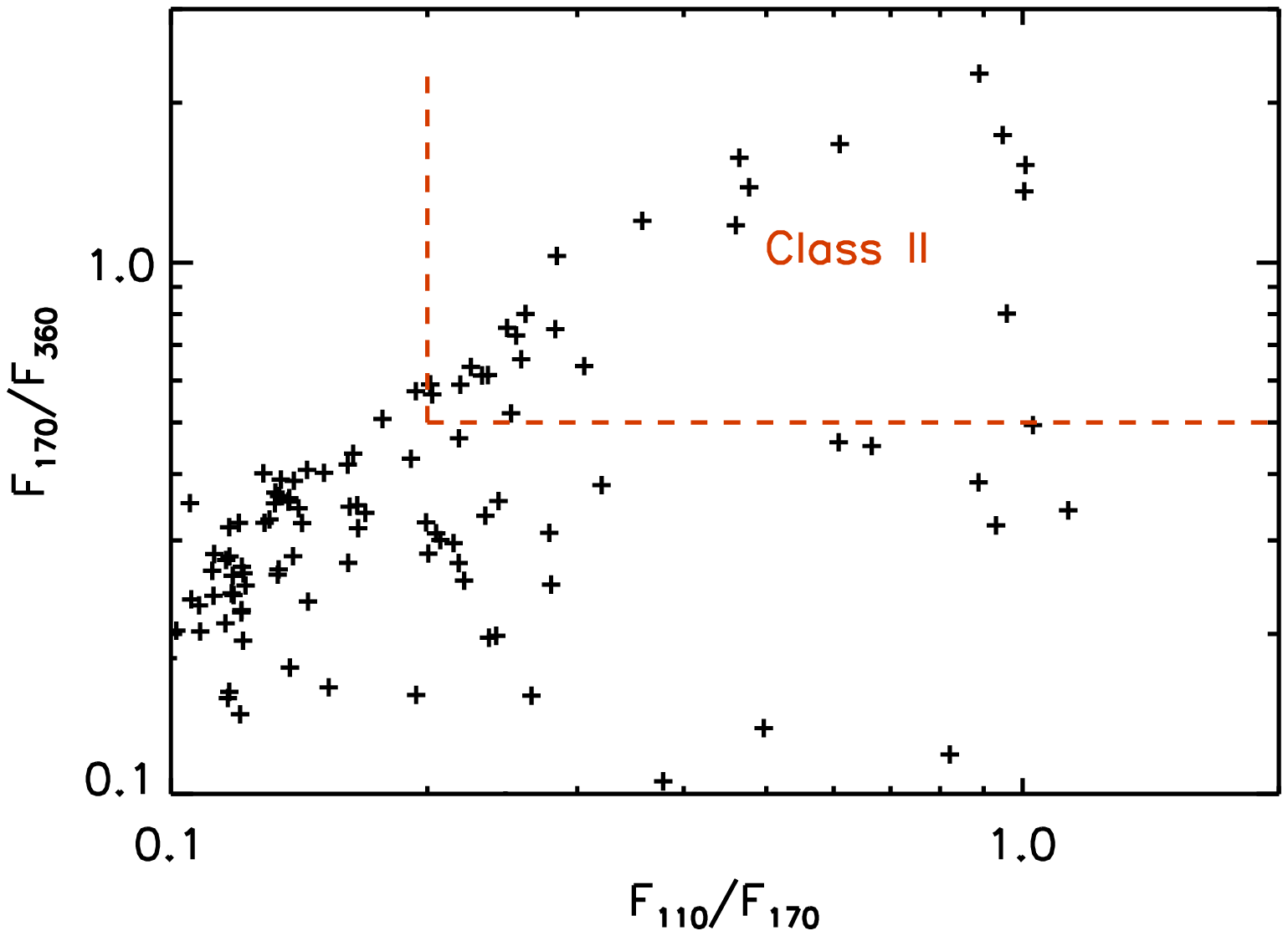,height=3.in,width=3.5in}}
\end{center}
\caption{Predictions for $\it{Herschel}$ color-color plots that depict the ``cold-warm'' trend (in observed frame) for $z=2$ galaxies with same lines as in previous two figures shown here averaged over viewing angles.}
\end{figure}

We see a striking trend when we plot these colors against each other -  most of the data is clustered at low $F_{75~\micron}/F_{170~\micron}$ and $F_{170~\micron}/F_{360~\micron}$ and a small fraction ($\la 15\%$) is in the energetically active AGN phase, i.e., the Class II phase, demarcated by the dashed lines in Figure 21.  As the earlier plots (Figures 19 and 20) made clear, galaxies tend to spend most of their time in low regions of $F_{75~\micron}/F_{170~\micron}$ and $F_{170~\micron}/F_{360~\micron}$ undergoing a brief excursion into high regions of the color-color plot, at which point the AGN dominates the energetics of the galaxy.  Although the correlation with either one of these colors with the AGN's relative luminosity is weakened by the dependence on viewing angle, high values of $\it{both~colors}$ (the upper right-hand quadrant in Figure 21) is firmly correlated with the AGN dominating the bolometric luminosity.  We have demarcated the energetically active AGN region by the dashed lines (upper right-hand quadrant) in a conservative way, i.e., we opt to identify galaxies as being in the AGN region when both colors always coincide with $L_{\rm BH} \ga 0.5 L_{\rm total}$.  In other words, we have left out sources which have $L_{\rm BH} \ga 0.5 L_{\rm total}$ but do not fall in our Class II region due to the dependence on viewing angle.   Figure 22 depicts the $F_{110~\micron}/F_{170~\micron}$ and $F_{170~\micron}/F_{360~\micron}$ colors for $z=2$ galaxies and shows a similar trend.  To emphasize the overall trend in the increase of particularly the $F_{110~\micron}/F_{170~\micron}$ and $F_{170~\micron}/F_{360~\micron}$ colors with the relative black hole luminosity, we plot these colors averaged over 200 viewing angles, with respect to the ratio of the black hole to the total bolometric luminosity in Figures 23.  These averaged colors plotted against each other in Figure 24 evince the same trend as shown in the complex Figure 22, i.e., galaxy colors clustering in low regions of $F_{110~\micron}/F_{170~\micron}$ and $F_{170~\micron}/F_{360~\micron}$, with a small fraction in the AGN region (upper right hand quadrant marked as ``Class II'').  We also show the cold-warm trend in Figure 25 for $z=3$ galaxies.  We have focused our attention on these epochs as it appears that multiple mergers become more prevalent at higher redshifts (Fakhouri \& Ma 2007), and fully cosmological simulations (Governato et al. 2007) may be needed to model the detailed time evolution of infrared bright galaxies at higher redshifts.  Moreover, the luminosity function of quasars peaks around $z \sim 2-3$ and drops sharply at higher epochs (Richards et al. 2006); as such, we have concentrated our attention here on the redshift range which is likely to yield the largest fraction of quasar hosts.    

\begin{figure}[!hb] \begin{center}
\centerline{\psfig{file=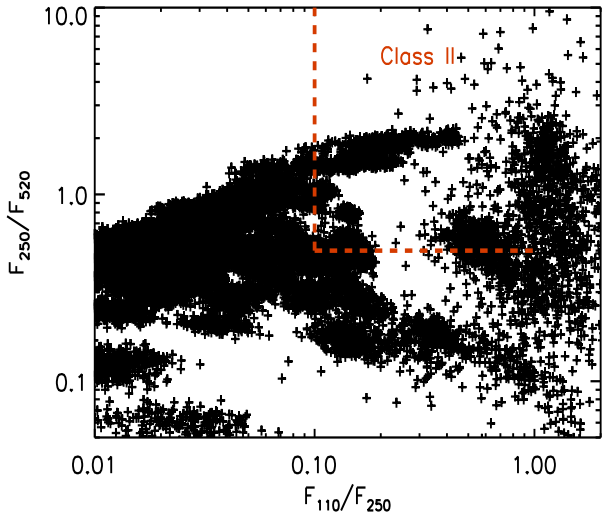,height=3.in,width=3.5in}}
\end{center}
\caption{Herschel color-color plots with the ``Class II'' evolutionary phase marked for $z=3$ galaxies.  Having both of these colors leads to a greater constraint for identifying energetically active AGN than one color alone.  Galaxies tend to cluster in low regions of $F_{\rm 110~\micron}/F_{\rm 250}$ and low $F_{\rm 250~\micron}/F_{\rm 520~\micron}$ during most of their lifetime, and spend a small fraction of their time in high regions of the $F_{\rm 110~\micron}/F_{\rm 250}$-$F_{\rm 250~\micron}/F_{\rm 520~\micron}$ color-color plot.  During this phase (when $\it{both}$ colors are high) the AGN contributes significantly to the bolometric luminosity and is effective at expelling obscuring columns of dust and gas, lowering the effective optical depth through the envelope - thereby producing ``warm'' colors.}
\end{figure}

Our results here suggest that secure identifications of energetically active AGN in the $z \sim 2-3$ SMG population on the basis of photometric colors would require $\it{wide-area}$ surveys.  Most infrared bright galaxies will cluster in the lower regions of $\it{Herschel}$ color-color plots, as galaxies spend most of their time as star forming galaxies, undergoing brief excursions into the energetically active AGN region (the upper right hand quadrant in Figures 21,22, 24 and 25 marked as ``Class II'').   

\section{Conclusion}

$\bullet$  We find that estimating the mean free intensity from the sum of photon flight paths and iteratively imposing radiative equilibrium (Lucy 1999) is a versatile and efficient temperature calculation algorithm, and particularly so for modeling three-dimensional dust envelopes.  We find that it is on average a factor of 10 times faster than relaxation algorithms for three-dimensional envelopes.  We have done extensive convergence studies to assess the viability of radiative equilibrium temperature algorithms, comparing to standard homogeneous test cases in one and three-dimensional grids.

$\bullet$ We discuss how many photons are needed to obtain a converged temperature profile and a high fidelity SED, which we have termed $N_{\rm temp}$ and $N_{\rm SED}$ respectively.  We find that the total number of photons required to get a converged temperature profile througout the entire dust envelope that is accurate to 2 \% can be accomplished with $N_{\rm temp,per~iter} \sim N_{\rm grid}$, where $N_{\rm temp,per~iter}$ is the number of photons per iteration and $N_{\rm grid}$ the number of grid cells.  Typically a total of four iterations are sufficient.  $N_{\rm SED} \sim 5 N_{\rm temp,per~iter}$ is sufficient to reach a noise-free long wave inclination averaged SED.  $N_{\rm SED} \sim 24 N_{\rm temp,per~iter}$ is generally sufficient to produce a low level of noise over viewing angles out to submillimeter wavelengths.

$\bullet$  We apply the code RADISHE to calculate the emergent SEDs of dusty galaxies along time sequences of hydrodynamical simulations.  Since RADISHE treats attenuation, scattering, and reemission self-consistently, it can generate a suite of accurate SEDs and corresponding dust temperatures calculated throughout the dust envelope.  We have compared our calculated SEDs during the Class I and Class II evolutionary phases to observed multi-wavelength data of SMGs to find good agreement.

$\bullet$  Performing radiative transfer calculations along a time sequence of SPH simulations allows us to construct simulated color-color plots, which we can unfold as a function of some intrinsic galaxy parameter, such as time or AGN luminosity.  We have previously used this decomposition to explain trends in IRAC color-color plots (Chakrabarti et al. 2007b).  Here, we predict that $\it{Herschel}$ color-color plots will be able to effectively discriminate between galactic energy sources in the $z \sim 2-3$ SMG population.  

$\bullet$ We have demarcated the energetically active AGN region - the Class II phase, in $\it{Herschel's}$ bands.  Specifically, we predict that (observed-frame) $F_{110~\micron}/F_{170~\micron}$-$F_{170~\micron}/F_{360~\micron}$ color-color plots will be able to most effectively select energetically active AGN in the $z \sim 2$ SMG population.  We emphasize the viewing angle dependence of these colors, and demarcate the Class II phase in a conservative manner.  Although the viewing angle dependence weakens the correlation in either of these colors with $L_{\rm BH}/L_{\rm total}$, the coincidence of $\it{both~colors}$ having high values is firmly correlated with $L_{\rm BH}/L_{\rm total} \ga 0.5$.  Since SMGs undergo a brief excursion into the Class II region, wide-area surveys will be needed to detect a statistically significant sample of $z \sim 2$ SMGs with energetically active AGN.

\bigskip
\bigskip

\acknowledgments

SC is indebted to Chris McKee for many insightful discussions on radiative transfer.  We also thank Leon Lucy, Jon Bjorkman, 
and George Rybicki for helpful discussions on temperature calculation algorithms, Giovanni Fazio, Jiasheng Huang, Matt Ashby, and Howard Smith for helpful feedback on infrared observations, T.J. Cox and Dusan Keres for informative discussions on SPH simulations, and Erik Rosolowsky for many helpful discussions on the properties of GMCs.  SC is supported by a NSF Postdoctoral Fellowship.  BW is supported by NASA Astrophysics Theory Program grant NNG 05-GH35G.  These calculations have been performed on the Institute for Theory and Computation Cluster.

\appendix


\section{I. 1-D Grids:  Fiducial Iterative Scheme}

Figure 26 shows the emergent SED from the homogeneous dust envelope considered in \S 3.1.1 when the Lucy method is used.  Four different exit criteria have been used ($V1-V4$) to end the iteration.  These are respectively:  $V1=\overline{T(r,\theta,\phi)_{i+1}-T(r,\theta,\phi)_{i}}$,  $V2=(\left|T^{rms}_{i+1}-T^{rms}_{i}\right|)/T^{rms}_{i+1}$, where $T^{rms}_{i}$ is defined as $\sqrt{\bar{T}_{i}^{2}+\sigma_{i}^{2}}$, where $\bar{T}_{i}$ is the average of the temperatures of all grid cells in iteration $i$.  $V3=\sqrt{1/N(\sum(T_{i}-\bar{T}))^{2}}$ where $\bar{T}$ is simply the average of the temperatures in all grid cells for that iteration. $V4$ is also a standard deviation, however, the average is taken with respect to the difference of temperatures in all grid cells in the $i+1$ iteration and the i'th iteration.  As is clear, these somewhat different ways of ending the iteration result in nearly identical emergent SEDs.  Figures 27 show the variation in these quantities with respect to iteration number.  The iteration is discontinued when the derivative of $V$ is no longer changing over iterations, at the 10 \% level (we show later that this is sufficient).  For the $V4$ method, typically four iterations are sufficient.  We adopt $V4$ as our fiducial iteration exit criterion since as a standard deviation, it incorporates information both about the average of and the relative spread in the temperatures per iteration.  It is also efficient, requiring 772.10 of CPU time on a 2 GHz AMD Opteron processor.  

\begin{figure}[!ht] \begin{center}
\centerline{\psfig{file=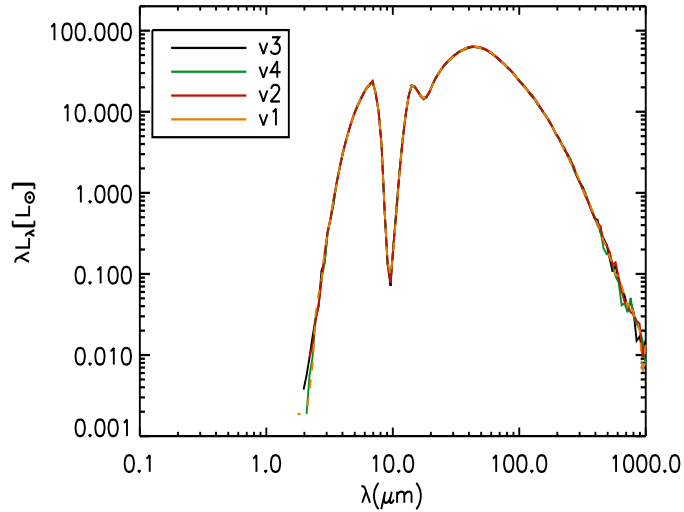,height=3.in,width=3.5in}}
\end{center}
\caption{Emergent SED from one-dimensional envelope for a spherically symmetric density profile using the iterative Lucy method with the four different exit criteria described in \S 2.}
\end{figure}

\begin{figure}[!ht] \begin{center}
\centerline{\psfig{file=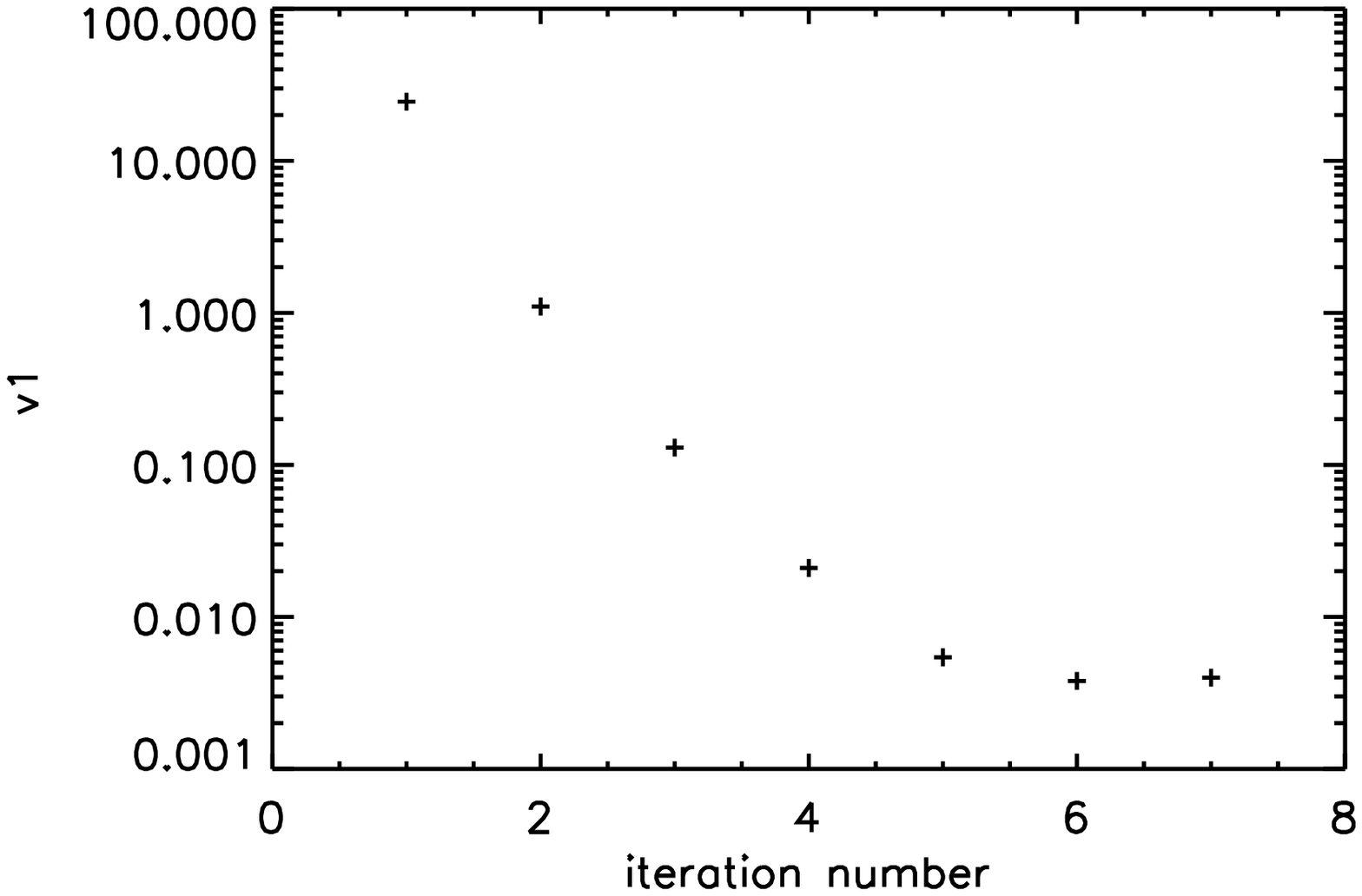,height=2.in,width=2.in}
\psfig{file=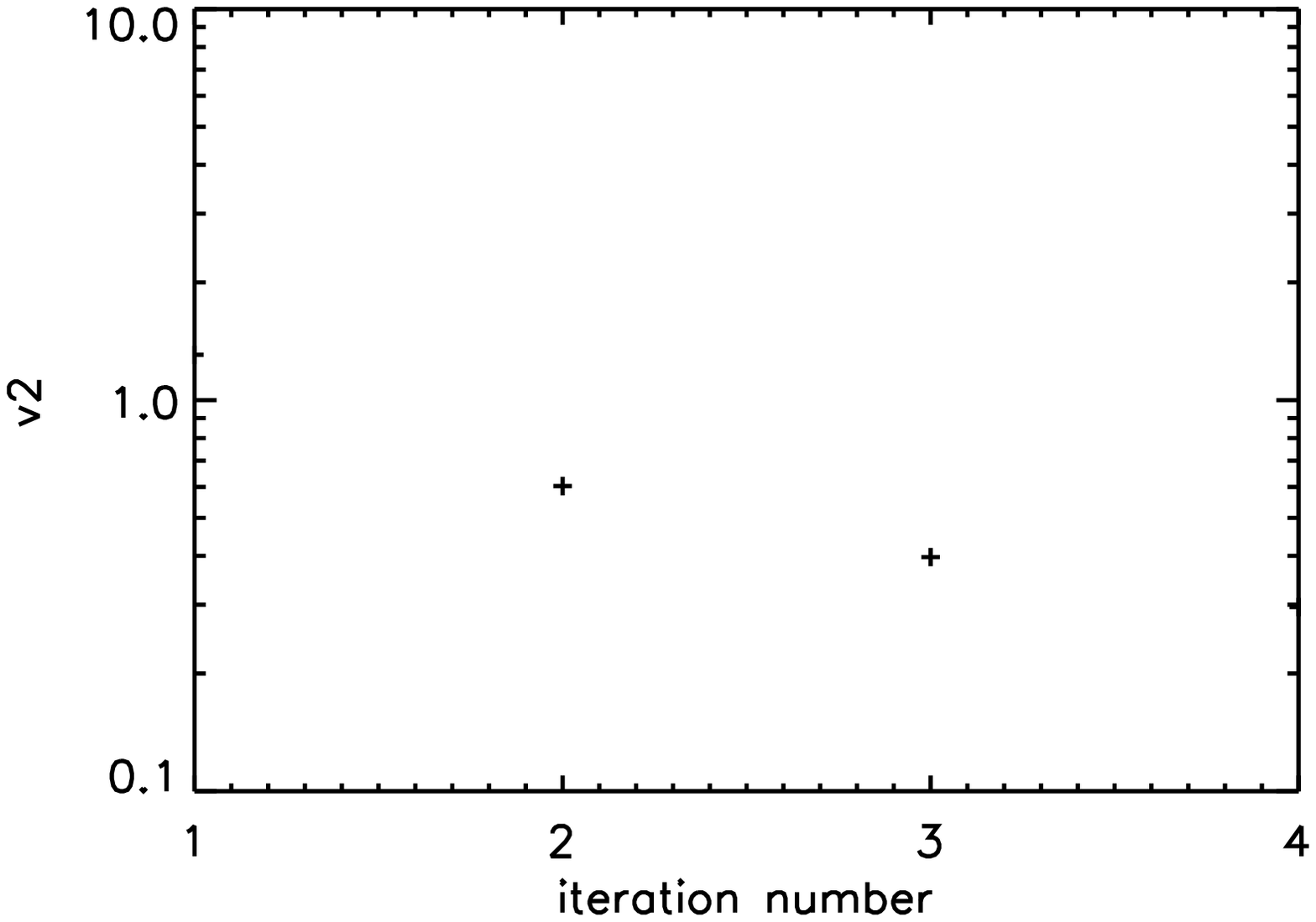,height=2.in,width=2.in}
\psfig{file=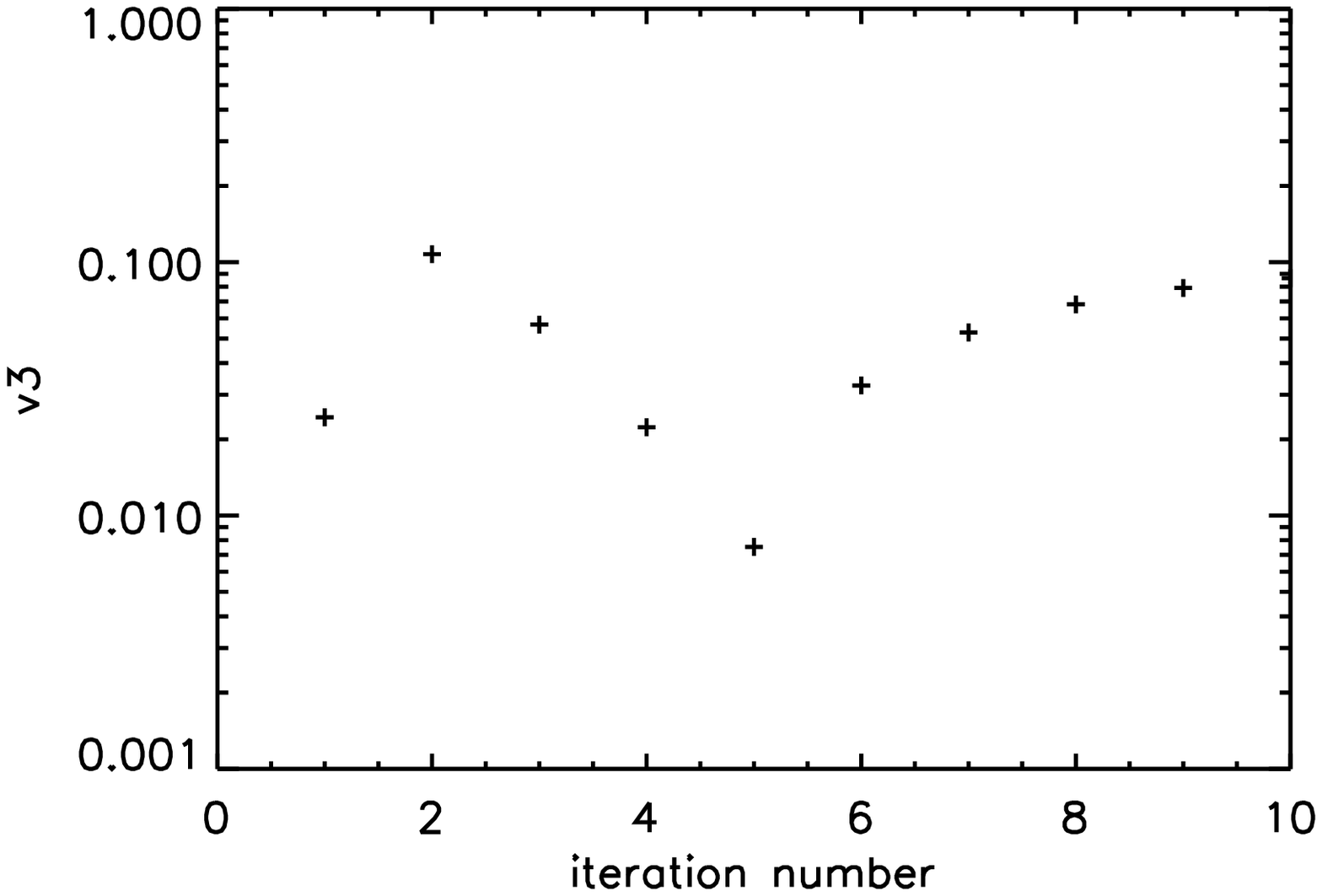,height=2.in,width=2.in}
{\psfig{file=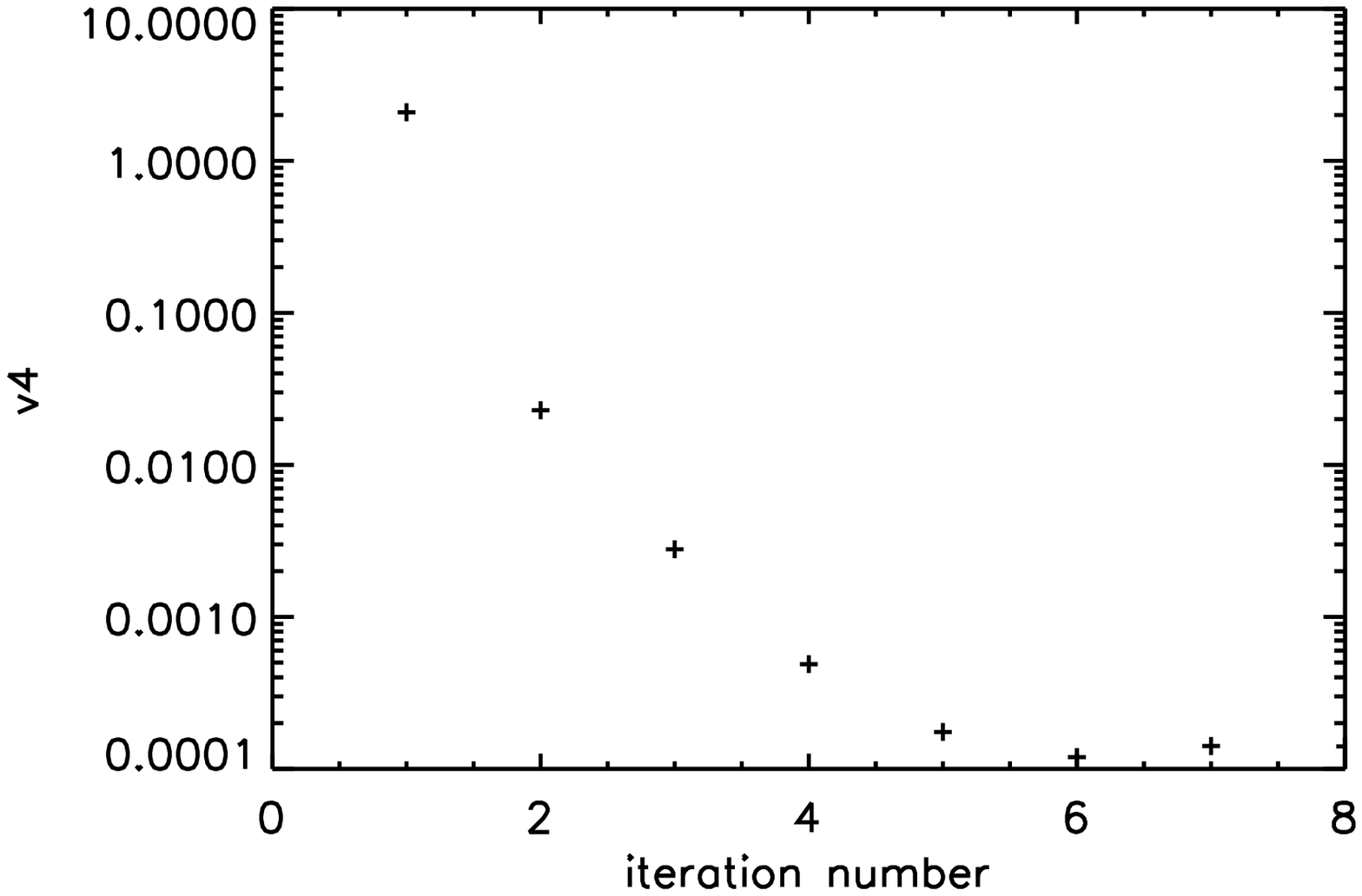,height=2.in,width=2.in}}}
\end{center}
\caption{Convergence when the $V1-V4$ (from \S 3) exit criteria are used to end the iteration of the temperature calculation - (a)convergence in the average temperature between the i'th and i+1 iteration, when the Lucy method is implemented to calculate the SED shown in Figure 3, (b)convergence in the root mean square of the temperature between the i'th and i+1 iteration, when the Lucy method is implemented to calculate the SED shown in Figure 3, (c)convergence in the standard deviation of the temperature, between the i'th and i+1 iteration, when the Lucy method is implemented to calculate the SED shown in Figure 3, (d)convergence in the standard deviation of temperature between the i'th and i+1 iteration, when the Lucy method is implemented to calculate the SED shown in Figure 26.}
\end{figure}

\section{II. 3-D Grids:  Standard Deviation Tolerance Levels \& $N_{\rm SED}$ for inclination averaged SEDs}

\begin{figure}[!ht] \begin{center}
\centerline{\psfig{file=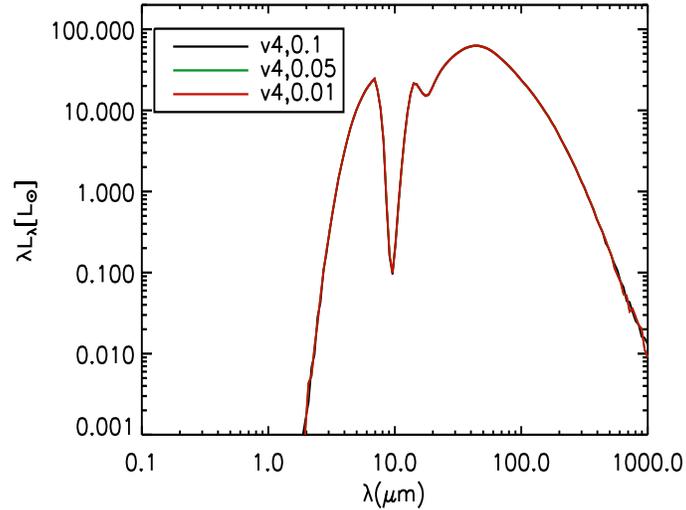,height=3.in,width=3.5in}}
\end{center}
\caption{Emergent SED from three-dimensional envelope (shown averaged over 200 viewing angles) for a spherically symmetric density profile using the iterative Lucy method and exiting by reference to the standard deviation (v4), for varying tolerance levels, i.e., when the standard deviation is below some number (0.1, 0.05, 0.1).  Here, each iteration is done with 6 million photons.}
\end{figure}

\begin{figure}[!ht] \begin{center}
\centerline{\psfig{file=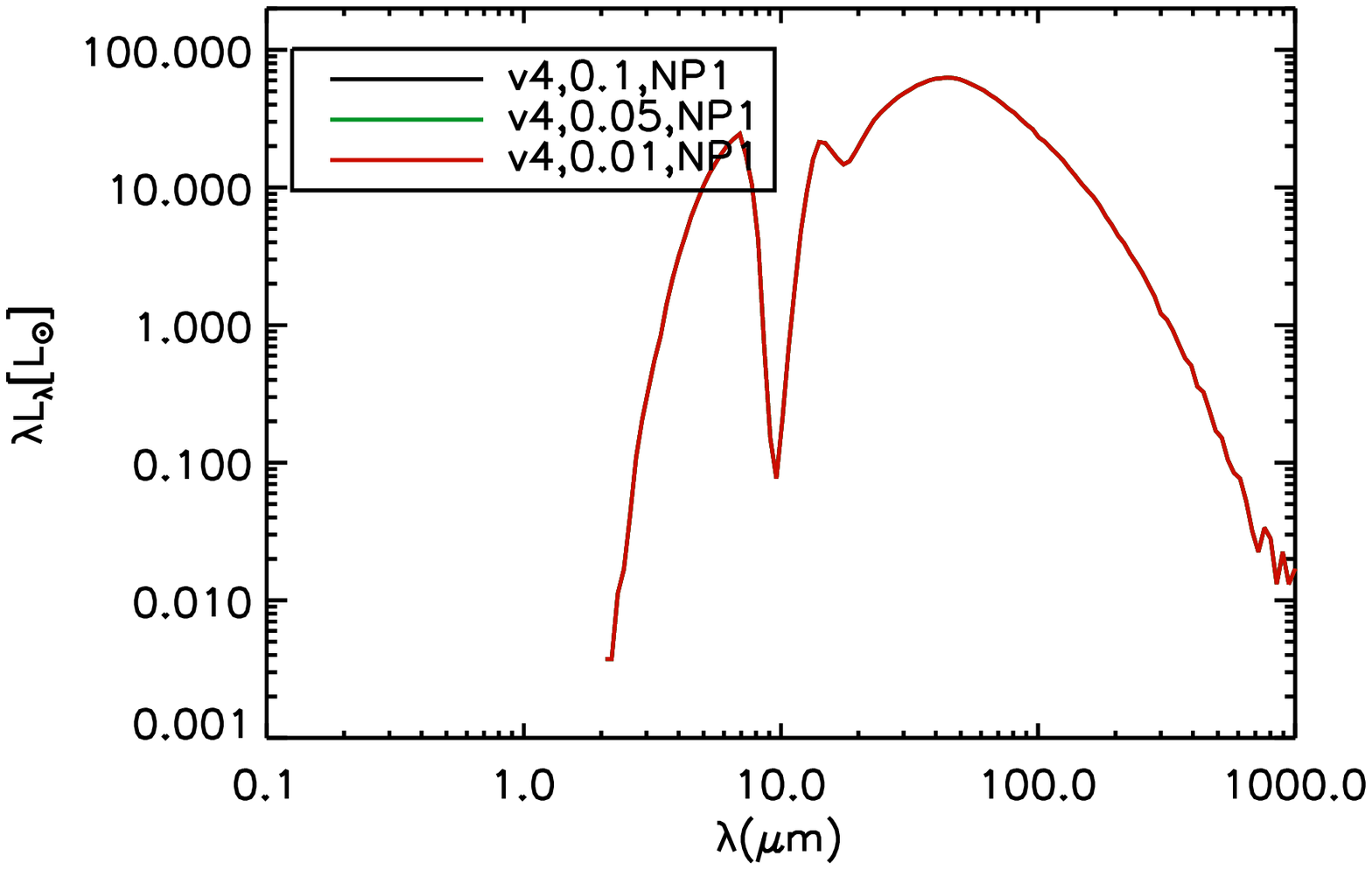,height=3.in,width=3.5in}}
\end{center}
\caption{Emergent SED from three-dimensional envelope (shown averaged over 200 viewing angles) for a spherically symmetric density profile using the iterative Lucy method and exiting by reference to the standard deviation (v4), for varying tolerance levels, i.e., when the standard deviation is below some number (0.1, 0.05, 0.1).  Here, each iteration is done with 1 million photons.  This is 6 times faster than iterating with 6 million photons per iteration but results in a noisier spectrum at the long wavelengths.}
\end{figure}

\begin{figure}[!ht] \begin{center}
\centerline{\psfig{file=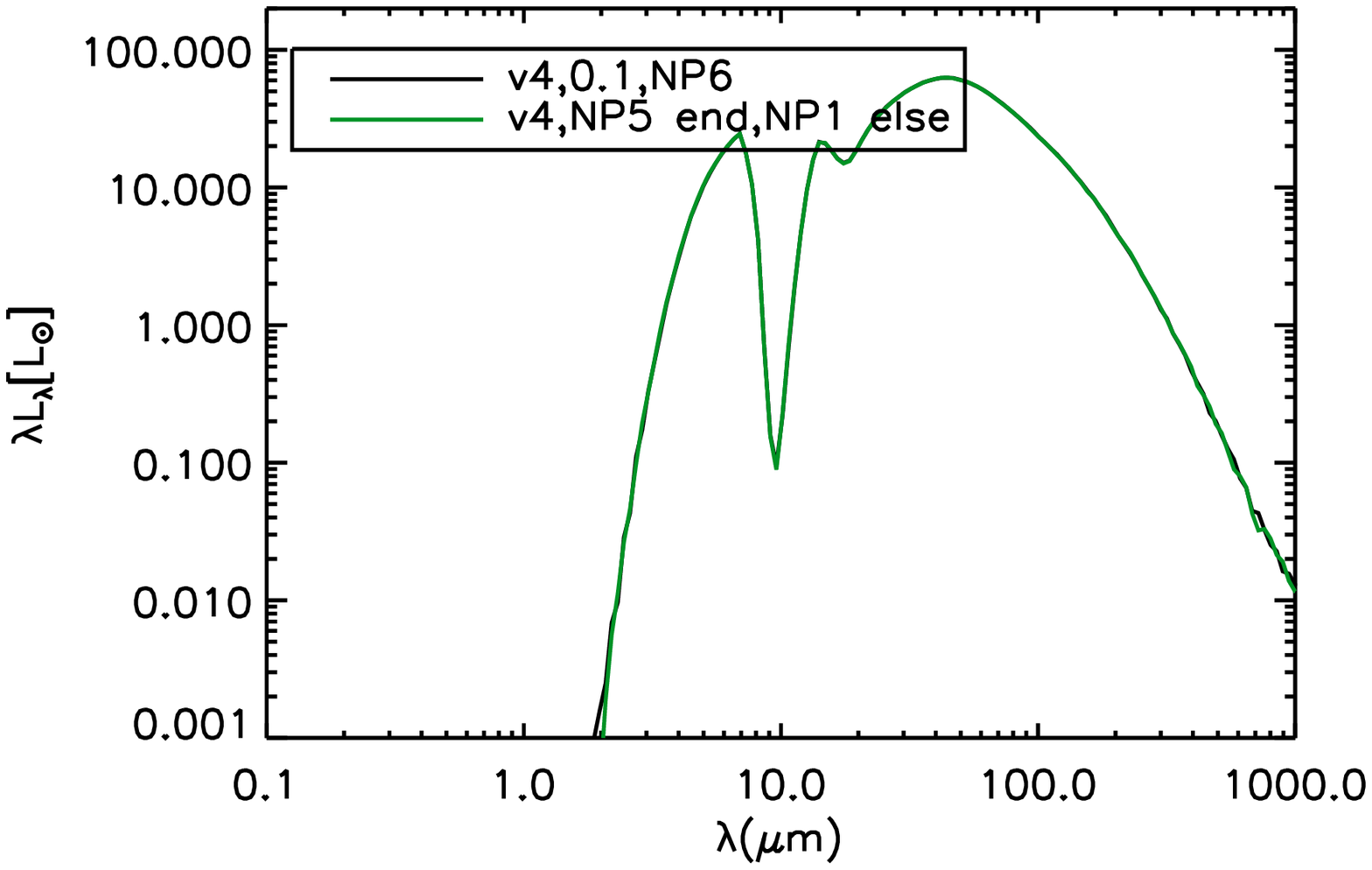,height=3.in,width=3.5in}}
\end{center}
\caption{Emergent SED from three-dimensional envelope (shown averaged over viewing angles) for a spherically symmetric density profile using the iterative Lucy method and exiting by reference to the standard deviation.  The iterations are done with 1 million photons each until the standard deviation falls below 0.1.  The last iteration is then done with 5 million photons.  This is faster than iterating with 6 million photons by almost a factor of 3, and is as accurate.  This is the fiducial iterative scheme to produce SEDs averaged over viewing angles}
\end{figure}

We first show the resultant SED using the Lucy method (exiting the iteration loop with respect to the standard deviation, $V4$) in Figure 28 when the iteration is continued to various tolerance levels for $V4$, i.e., $V4<0.1,0.05,0.01$.  
Here, each iteration has been done with $6 \times 10^{6}$ photons.  As is clear, iterating past a tolerance of 0.1 does not change the results.  Iterating to a tolerance of 0.1 took 7896 s of CPU time.  Successive iterations, i.e., to reach 0.05 or 0.01 tolerance are relatively fast, requiring 10,000s and 10,094s respectively.  We next show in Figure 29 the SEDs when $1\times 10^{6}$ photons have been used per iteration.  This results in a noisier SED at the longer wavelengths even when the tolerance is low, but is about six times faster than iterating with $6\times 10^{6}$ photons.  Iterating to 0.1 tolerance with $1\times 10^{6}$ photons required only 1382s of CPU time.  Since iterating with fewer photons is much faster, we choose to iterate with $1\times 10^{6}$ photons until the standard deviation is changing at less than 0.1 with respect to iteration number.  We then iterate one more time with $5\times 10^{6}$ photons.  This case is shown in Figure 30.  It yields a converged, essentially noise-free SED quickly in 3326s of CPU time.  Due to its efficiency, we adopt this as our fiducial iterative scheme.

\end{document}